\documentclass[
 aip,
 jcp,
 amsmath,amssymb,
reprint,
]{revtex4-2}

\usepackage{graphicx}
\usepackage{dcolumn}
\usepackage{bm}

\usepackage[utf8]{inputenc}
\usepackage[T1]{fontenc}
\usepackage{mathptmx}
\usepackage{etoolbox}
\usepackage{mathtools}

\newcommand{\abs}[1]{{\vert #1\vert}}

\newcommand{\avg}[1]{{\langle #1\rangle}}
\newcommand{\zc}[0]{{\mathcal{Z}_\text{c}}}
\newcommand{\zgc}[0]{{\mathcal{Z}_\text{gc}}}
\newcommand{\at}[2]{{\left. #1\right\vert_{#2}}}

\makeatletter
\def\@email#1#2{%
 \endgroup
 \patchcmd{\titleblock@produce}
  {\frontmatter@RRAPformat}
  {\frontmatter@RRAPformat{\produce@RRAP{*#1\href{mailto:#2}{#2}}}\frontmatter@RRAPformat}
  {}{}
}%
\makeatother

\begin{document}

\title{Variational functional theory for coulombic correlations in the electric double layer}

\author{Nils Bruch}
\affiliation{Theory and Computation of Energy Materials (IET-3), Institute of Energy Technologies, Forschungszentrum J\"ulich GmbH, 52425 J\"ulich, Germany}
\affiliation{Chair of Theory and Computation of Energy Materials, Faculty of Georesources and Materials Engineering, RWTH Aachen University, 52062 Aachen, Germany}

\author{Tobias Binninger}
\email{t.binninger@fz-juelich.de}
\affiliation{Theory and Computation of Energy Materials (IET-3), Institute of Energy Technologies, Forschungszentrum J\"ulich GmbH, 52425 J\"ulich, Germany}

\author{Jun Huang}
\affiliation{Theory and Computation of Energy Materials (IET-3), Institute of Energy Technologies, Forschungszentrum J\"ulich GmbH, 52425 J\"ulich, Germany}
\affiliation{Chair of Theory and Computation of Energy Materials, Faculty of Georesources and Materials Engineering, RWTH Aachen University, 52062 Aachen, Germany}

\author{Michael Eikerling}
\affiliation{Theory and Computation of Energy Materials (IET-3), Institute of Energy Technologies, Forschungszentrum J\"ulich GmbH, 52425 J\"ulich, Germany}
\affiliation{Chair of Theory and Computation of Energy Materials, Faculty of Georesources and Materials Engineering, RWTH Aachen University, 52062 Aachen, Germany}

\begin{abstract}
	A classical coulombic correlation functional in one-loop (1L) and local-density-approximation (LDA) is derived for electrolyte solutions, starting from a first-principles many-body partition function. The 1L-LDA functional captures correlations between electrolyte ions and solvent dipoles, such as screening and solvation, that are ignored by conventional mean-field theories. 
    This 1L-LDA functional introduces two parameters that can be tuned to the experimental dielectric permittivity and activity coefficients in the bulk electrolyte solution. 
    The capabilities of the 1L-LDA functional for the description of metal--electrolyte interfaces are demonstrated by embedding the functional into a combined quantum--classical model. Here, the 1L-LDA functional leads to a more pronounced double-peak structure of the interfacial capacitance with higher peaks and shorter peak-to-peak distance, significantly improving the agreement with experimental data and showing that electrolyte correlation effects exert a vital impact on the capacitive response.
\end{abstract}

\maketitle

\section{Introduction}
Defossilization of the energy sector requires cost-efficient electrolyzers and fuel cells that are capable of converting electrical energy into chemical energy in the form of hydrogen and back into electrical energy.\cite{galloEnergyStorageEnergy2016} Enhancing current technologies necessitates a deeper understanding of the microscopic region between solid metal and liquid electrolyte, in which electrocatalytic reactions occur. The interfacial region between a charged metal surface and an electrolyte solution has been a topic of long-standing interest in various scientific fields, encompassing electrochemistry and biology.\cite{fendlerColloidalDomainWhere1996, mclaughlinElectrostaticPropertiesMembranes1989} However, the structure of the electric double layer (EDL), including electric field and density distributions of electrolyte species, solvent dipole orientation, and dielectric permittivity, is challenging to probe experimentally.\cite{schwarzElectrochemicalInterfaceFirstprinciples2020} Therefore, theoretical approaches are crucial for complementing experimental insights.

\textit{Ab initio} molecular dynamics (AIMD) simulations based on quantum-mechanical density functional theory (DFT) can be used to simulate metal and electrolyte at an atomistic level.\cite{leModelingElectrochemicalInterfaces2020} However, such AIMD approaches are still computationally infeasible for complex realistic systems.\cite{grossChallengesInitioMolecular2023} To address the challenge of computational cost, various hybridization schemes have been developed that describe electrode and electrolyte regions at different levels of theory.\cite{sundararamanGrandCanonicalElectronic2017} The metal description varies from Kohn-Sham DFT, as used in ESM-RISM\cite{nishiharaHybridSolvationModels2017,teschPropertiesPt1112021,hagiwaraDevelopmentDielectricallyConsistent2022,schmeerDevelopmentThermodynamicProperties2010,gusarovSelfConsistentCombinationThreeDimensional2006}  or joint density-functional theory (JDFT)\cite{petrosyanJointDensityfunctionalTheory2007,sundararamanComputationallyEfficaciousFreeenergy2012,melanderGrandcanonicalApproachDensity2019}, to orbital free DFT (OFDFT) as employed in density-potential functional theory (DPFT)\cite{huangDensityPotentialFunctionalTheory2023,shibataParameterFittingFreeContinuumModeling2024} and classical electrode models with fluctuating atomic charges.\cite{ corettiMetalWallsSimulatingElectrochemical2022} On the other hand, the electrolyte description varies from classical molecular dynamics (MD) simulations for sampling the spatial movement of electrolyte species\cite{dohmDevelopingAdaptiveQM2017,abidiElectrostaticallyEmbeddedQM2023} to continuum electrolyte models. Significant efforts are devoted to improving the continuum description of electrolyte solutions that are able to capture the essential physics of the EDL.

Continuum approaches, where the picture of individual particles is smeared out to continuous (averaged) density distributions, were the basis for the earliest EDL models. In the Helmholtz model\cite{helmholtzUeberEinigeGesetze1853} the EDL is described by a simple capacitor while in the Gouy-Chapman-Stern (GCS)\cite{sternZurTheorieElektrolytischen1924} approach, the EDL is further refined by an additional \emph{diffuse} layer, with the mean-field (MF) Poisson-Boltzmann (PB) equation employed to calculate potential and density distributions. However, the GCS model does not quantitatively agree with experimental capacitance data for larger electrode potentials.\cite{hatloElectricDoubleLayer2012} These shortcomings of the GCS model are addressed in augmented PB approaches, such as the Modified PB (MPB) model that considers sterical, or finite-size, effects\cite{borukhovStericEffectsElectrolytes1997,bikermanStructureCapacityElectrical1942} to improve the agreement with experimental capacitance data.\cite{bazantUnderstandingInducedchargeElectrokinetics2009}

In electrostatic MF models, such as the GCS theory, particles interact solely through an average electrostatic potential generated by the average charge distribution of all particles, which means that electrostatic correlation effects are not accounted for. Coulombic correlations are however crucial for various properties of electrolytes. Ion--solvent correlations induce the typical dielectric decrement,\cite{adarDielectricConstantIonic2018} while ion--ion correlations reduce the ionic activity coefficient due to screening, as described in Debye-Hückel theory.\cite{Debye_1923} A variety of models have been proposed to include coulombic correlation effects, such as over-screening\cite{bazantDoubleLayerIonic2011} or excess ion-polarizability\cite{hatloElectricDoubleLayer2012}, into the GCS theory \textit{ad hoc}, which was shown to provide quantitative agreement with experiments. 

Classical DFT (cDFT) provides a versatile theoretical framework for a systematic treatment of electrolyte solutions. In cDFT, one seeks an expression for a free energy functional, which is minimized by the equilibrium density profiles.\cite{hansenTheorySimpleLiquids2006} The simplest functional capturing coulombic interaction is the MF functional, which neglects coulombic correlation effects and reduces to the classical PB equation when minimized.\cite{bultmannPrimitiveModelClassical2022} Interactions beyond MF are described by a correlation functional. However, as in quantum-mechanical DFT, the exact density functional for electrolytes is unknown.\cite{evansNatureLiquidvapourInterface1979} In practice, the formally exact Ornstein-Zernike (OZ) integral equation, which relates the direct correlation function to the (experimentally measurable) pair correlation function, is employed for constructing approximate cDFT correlation functionals.\cite{hansenTheorySimpleLiquids2006} The OZ equation needs a closure relation, such as the mean spherical approximation (MSA), from which functionals have been derived.\cite{kierlikDensityfunctionalTheoryInhomogeneous1991,gillespieCouplingPoissonNernst2002,rothShellsChargeDensity2016} These functionals capture important electrolyte phemonena such as screening\cite{rothShellsChargeDensity2016,ebelingEstimationTheoreticalIndividual1983} but the approximations made by closure relations are ambiguous and difficult to improve.\cite{bultmannPrimitiveModelClassical2022} Another limitation is the implicit treatment of the solvent, meaning that while ion--ion correlations are considered through the OZ equation, solvent--solvent and ion--solvent correlations are typically not included. In molecular density functional theory (MDFT),\cite{ramirezDensityFunctionalTheory2002,jeanmairetMolecularDensityFunctional2013} the solvent has been explicitly incorporated in the OZ equation, but this approach requires additional molecular dynamics (MD) simulations for the computation of the pair correlation function. 

Alternatively, electrolyte solutions have been treated by a field theoretic approach that does not rely on the OZ integral equation.\cite{netzPoissonBoltzmannFluctuationEffects2000} The statistical partition function of an electrolyte is mapped by a Hubbard-Stratonovic (HS) transformation\cite{hubbardCalculationPartitionFunctions1959,stratonovichMethodCalculatingQuantum1957} to an exact functional integral of the electrostatic potential weighted by an action functional. This form is very attractive as it resembles a functional integrals familiar from quantum field theory (QFT) that have been extensively studied in the past century.\cite{altlandCondensedMatterField2010,zinn-justinQuantumFieldTheory2021} The toolbox of QFT allows systematic perturbative studies of thermodynamic quantities.\cite{podgornikAnalyticTreatmentFirstorder1990} Pioneering works by Netz and Orland reproduced Debye Hückel (DH) results for bulk systems using this field theoretic approach.\cite{netzDebyeHUckelTheory1999} Subsequent extensions of the field theoretic framework, derived with the one-loop (1L) approximation for the electrostatic potential,\cite{netzPoissonBoltzmannFluctuationEffects2000} elucidated various phenomena such as dielectric decrement in bulk electrolytes,\cite{adarDielectricConstantIonic2018} coulombic correlations at dielectric interfaces,\cite{markovichSurfaceTensionElectrolyte2015, santangeloComputingCounterionDensities2006} and non-local solvent structure.\cite{desouzaPolarLiquidsCharged2022, blosseyFieldTheoryStructured2022} Combining the toolbox of the field theoretic approach with the cDFT framework appears highly promising but, to the best of our knowledge, has not been presented yet.

This article presents two main results. Firstly, in Sec.~\ref{Sec: Theory}, it is shown via formal derivation that the field-theoretic approach leads to the definition of a correlation functional for an electrolyte solution. By applying the 1L approximation together with a local-density approximation (LDA), we derive a 1L-LDA correlation functional, \(\mathcal{F}^\text{1L-LDA}_\text{sol}\) (Eq.~\ref{CorrelationFunctional}), as the first-order correction to the coulombic MF functional. Readers who are more interested in the physical implications of using \(\mathcal{F}^\text{1L-LDA}_\text{sol}\) may choose to skip Sections \ref{Sec: Theory} and \ref{Sec: Application}. Secondly, in Sec.~\ref{SubSec:Bulk}, it is demonstrated that \(\mathcal{F}^\text{1L-LDA}_\text{sol}\) captures dielectric decrement and variation of the activity coefficient with ion concentration in bulk  electrolyte solutions. Embedding the derived functional into an EDL model (Sec.~\ref{SubSec:Interface}) achieves quantitative agreement with experimental capacitance data, which can be attributed to the inclusion of coulombic correlation effects. The reduction of the local interface activity coefficient of the electrolyte, leading to increased counterion density, results in more pronounced capacitance peaks with smaller peak-to-peak distances. This finding underscores the relevance of coulombic correlation effects for accurately describing the electric double layer (EDL) even in dilute electrolyte solutions.

\section{Theory\label{Sec: Theory}}

The objective of this section is to derive the 1L-LDA correlation functional for electrolyte solutions from first principles, utilizing a statistical mechanics approach. Starting with the potential energy of the electrolyte solution, the system's partition function can be expressed. Through a Hubbard-Stratonovic transformation, a functional integral representation is then obtained.Sec.~\ref{SubSec:TransitionVariational} presents a general derivation of a variational functional for the electrostatic potential and electrolyte solution densities. Sec.~\ref{SubSec:OneLoopAndLocalDensity} utilizes the insights from Sec.~\ref{SubSec:TransitionVariational} to derive an explicit variational functional for electrolyte solutions. This functional comprises the ideal gas functional, as described in Eq.~\eqref{idealGasFreeEnergy}, the excess MF functional, Eq.~\eqref{MeanFieldExcessFreeEnergy}, and crucially, the newly derived correlation functional in LDA, presented in equation \eqref{CorrelationFunctional}, which represents the key scientific result of this paper. This allows investigating the influence of coulombic correlations on the EDL in Sec.~\ref{Sec: Application}.

\subsection{Hubbard-Stratonovic transformation}
\label{SubSec:Hubbard-Stratonovic}
Following references  \onlinecite{netzPoissonBoltzmannFluctuationEffects2000,abrashkinDipolarPoissonBoltzmannEquation2007}, the potential energy of an electrolyte solution under the influence of spatially dependent external potentials $v_j^\text{ext}(r)$ is given by
\begin{align}
    U=\frac{1}{2}\int_{r,r'}\rho(r)V(r,r')\rho(r') + \int_r \sum_{j=a/c/s} v_j^\text{ext}(r)n_j(r),
\end{align}
where $\int_r$ is a shorthand notation for $\int d^3r$ and $r$ is a three-dimensional vector. The external potentials will be needed to describe the interaction with the electrode subsystem in Sec.~\ref{Sec: Application}.  The first term entails the Coulomb interaction,
\begin{align}
V(r,r')=\frac{1}{4\pi \epsilon_0 \vert r-r'\vert},
\end{align}
with vacuum permittivity $\epsilon_0$ and electrolyte charge density $\rho(r)$, while the second term is the linear coupling of $v_j^\text{ext}(r)$ to the local densities $n_j(r)$ of anions (a), cations (c) and solvent (s), respectively. The electrolyte consists of $N_{a}$ point-like anions with charge $q_a$ and $N_{c}$ cations with charge $q_c$, dissolved in $N_{s}$ solvent molecules with dipole moment $\vec{p}$, giving a total charge density of
\begin{align}\label{ChargeDensity}
\rho(r)=\sum_{i=a/c}q_{i}n_{i}(r)+\rho_{s}(r),
\end{align}
where the density of electrolyte particles is 
\begin{align}\label{DensityOperator}
n_{j}(r)=\sum_{k=1}^{N_{j}}\delta(r-r_{j,k}), \quad j=a,c,s.
\end{align}
and the dipole charge density\cite{abrashkinDipolarPoissonBoltzmannEquation2007} is 
\begin{align}\label{DipoleChargeDensity}
\rho_{s}(r)=\sum_{k=1}^{N_{s}}(\vec{p}_{k}\cdot\nabla)\delta(r-r_{s,k}).
\end{align}
Note that the dipole charge density is not identical to the dipole density. The dipole-induced charge density is the divergence of the dipole-induced polarization, e.g., Eq.~\eqref{DipoleChargeDensity}, while the dipole density is given by Eq.~\eqref{DensityOperator}.  

The fundamental quantity of a grand canonical system in equilibrium is the grand potential, 
\begin{align}\label{GrandPotentialFunction}
\Omega_\text{sol}(\mu_{j})=-\frac{1}{\beta}\log\mathcal{Z}_{\text{gc}}(\mu_{j}),
\end{align}
which is a function of the chemical potentials $\mu_j$ of the particle species $j$. The grand canonical partition function,
\begin{align}\label{Defn:PartitionFunction}
    \mathcal{Z}_{\text{gc}}(\mu_{j})=\sum_{N_{a}}\sum_{N_{c}}\sum_{N_{s}}\lambda^{N_{a}}\lambda^{N_{c}}\lambda^{N_{s}}\cdot\mathcal{Z}_{c}(N_{a},N_{c},N_{s}),
\end{align}
where $\lambda_j=\exp(\beta\mu_j)$ is the fugacity, can be obtained from the canonical partition function $\zc$, which is the phase-space integral over all possible electrolyte configurations weighted by their respective Boltzmann factors,
\begin{align}\label{Canonical Partition Function}
\mathcal{Z}_c(N_{j})=&\prod_{i=a/c}\frac{1}{N_{i}!\Lambda_{i}^{3N_i}}\prod_{k=1}^{N_{i}}\int_{r_{i,k}} \nonumber\\
&\frac{1}{N_{s}!\Lambda_{s}^{3N_s}}\prod_{k=1}^{N_{s}}\int_{r_{s,k}}\int_{d\Omega_{k}} 
\exp\left(-\beta U\right),
\end{align}
where $\beta=1/(k_BT)$ is the inverse temperature and $k_B$ is the Boltzmann constant. The partition function is written as an integral over all possible particle degrees of freedom. In this case, degrees of freedom include the positions $r_{j,k}$ of particles $k$ of type $j$ and the orientations $d\Omega_{k}$ of dipoles of solvent molecules. The factorial $N_{j}!$ takes care of overcounting, while integration over momentum degrees of freedom yields the thermal wavelengths $\Lambda_{j}^{3N_j}$. In general, this partition sum is not amenable to being transformed into an exact solution. However, in order to use approximation methods, following the work by Podgornik,\cite{podgornikAnalyticTreatmentFirstorder1990}  we rewrite the partition function as a functional integral over an auxiliary field linearly coupled to the charge densities. This can be achieved by Hubbard-Stratonovic (HS) transformation.\cite{hubbardCalculationPartitionFunctions1959} Starting from the Boltzmann factor, Eq.~\eqref{Canonical Partition Function}, we write
\begin{align}\label{GaussianIntegral}
\exp&\left(-\frac{\beta}{2}\int_{r,r'}\rho(r)V(r,r')\rho(r')\right) \nonumber \\
=\int &D\psi\cdot\exp\bigg(-\frac{\beta}{2}\int_{r,r'}\psi(r)V^{-1}(r,r')\psi(r') \nonumber \\
&-i\beta\int_{r}\psi(r)\cdot\rho(r) \bigg),
\end{align}
where we have neglected an irrelevant constant, $\sim\det(V^{-1})^{1/2}$. This allows the canonical partition function to be expressed as
\begin{align}\label{PositionIntegrals}
\mathcal{Z}_c(N_{j})	=	\int D\psi\exp\bigg(&-\frac{\beta}{2}\int_{r,r'}\psi(r)V^{-1}(r,r')\psi(r')\bigg) \nonumber \\
\prod_{i=a/c}\frac{1}{N_{i}!\Lambda_{i}^{3 N_i}}\prod_{k=1}^{N_{i}}\int_{r_{i,k}}&\frac{1}{N_{s}!\Lambda_{s}^{3N_k}}\prod_{k=1}^{N_{s}}\int_{r_{s,k}}\int_{d\Omega_{k}} \nonumber \\
\exp\Bigg(-i\beta\int_{r}\psi(r)&\rho(r) \nonumber\\-\beta\int_r \sum_{j=a/c/s} &v_j^\text{ext}(r)n_j(r)\Bigg).
\end{align}
The rewritten form makes it possible to simplify the position and orientation integrals, since charge density appears only linearly in the exponent. Note that the HS field $\psi$ is strictly real. After inserting the explicit form of the charge density (Eq.~\eqref{ChargeDensity}) into Eq.~\eqref{PositionIntegrals}, the space integration in the exponent can be performed to yield
\begin{align}
&\exp\bigg(-i\beta\big(\sum_{i,k}^{N_i} q_i \psi(r_{i,k}) + \sum_k^{N_s} \vec{p}_k\cdot \nabla \psi(r_{s,k}) \big)\bigg) \nonumber\\
&\times\exp\bigg(+\beta\sum_{j,k}^{N_j} v_j^\text{ext}(r_{j,k})\bigg).
\end{align}
The terms in the product over $N_i$ and $N_s$ can be rearranged to
\begin{align}
&\prod_{i=a/c} \prod_{k=1}^{N_i}  \frac{1}{\Lambda_i^{3}}\int_{r_{i,k}} \exp(-i\beta \sum_{i,k}^{N_i} q_i \psi(r_{i,k})-\beta\sum_{i,k}^{N_i} v^{\text{ext}}_i(r_{i,k}))  \nonumber \\
&=\prod_{i=a/c} \bigg(\int_{r_i} \Lambda_i^{-3} \exp(-i\beta q_i \psi(r_{i})-\beta v_{i}^\text{ext}(r_i))\bigg)^{N_i}, \\
&\prod_{k=1}^{N_s} \frac{1}{\Lambda_s^{3}} \int_{r_k} \int_{d\Omega_{k}} \exp(-i\beta \sum_{k}^{N_s} \vec{p}_k\cdot \nabla\psi(r_{i,k})-\beta\sum_{k}^{N_s} v_s^\text{ext}(r_{s,k}))  \nonumber\\
&=\bigg(\int_r \int_{d\Omega} \Lambda_s^{-3}\exp(-i\beta \vec{p}\cdot\nabla\psi(r)-\beta v_s^\text{ext}(r))\bigg)^{N_s}. \label{orientational integral}
\end{align}
Furthermore, the orientational integral, $d\Omega_{k}$, in Eq.~\eqref{orientational integral} can be evaluated by fixing the dipole strength to $p$ and aligning the dipole vector $\vec{p}$ with the $z$ axis. In this case, the scalar product in Eq.~(\ref{orientational integral}) is written as $p \left\vert\nabla \psi(r) \right\vert \cos(\theta)$, where $\theta$ is the angle between $\vec{p}$ and $\nabla\psi(r)$. Performing the integration then gives
\begin{align}\label{ResultCanonicalPartitionFunction}
&\mathcal{Z}_c(N_{j})	=	\int D\psi \nonumber\\
&\exp\bigg(-\frac{\beta}{2}\int_{r,r'}\psi(r)V^{-1}(r,r')\psi(r')\bigg) \nonumber \\
&\prod_{i=a/c}\frac{1}{N_{i}!}  \bigg(\int_{r_i} \Lambda_{i}^{-3} \exp(-i\beta q_i \psi(r_{i})-\beta v_{i}^\text{ext}(r_i))\bigg)^{N_i} \nonumber\\
&\frac{1}{N_{s}!} \bigg(\int_r \Lambda_{s}^{-3}e^{-\beta v_s^\text{ext}(r)}\frac{\sinh(ip\beta \vert \nabla\psi(r) \vert)}{ip\beta \vert \nabla\psi(r)\vert}\bigg)^{N_s}. 
\end{align}
Inserting Eq.~\eqref{ResultCanonicalPartitionFunction} into Eq.~\eqref{GrandPotentialFunction} and identifying the sum over particle numbers as the series expansion of the exponential function, one gets
\begin{align}
&\mathcal{Z}_{\text{gc}}(\mu_{j})=\sum_{N_{a}}\sum_{N_{c}}\sum_{N_{s}}\lambda_a^{N_{a}}\lambda_c^{N_{c}}\lambda_s^{N_{s}}\cdot\mathcal{Z}_{c}[N_{a},N_{c},N_{s}] \nonumber\\
&=\int D\psi\nonumber \\
&\exp\left(-\frac{\beta}{2}\int_{r,r'}\psi(r)V^{-1}(r,r')\psi(r')\right) \nonumber \\
&\exp\bigg( \int_r \sum_{i=a/c}\lambda_i\Lambda_{i}^{-3}e^{-\beta v_{i}^\text{ext}(r)} \exp(-i\beta q_i \psi(r))\nonumber  \\
&+\int_r \lambda_s\Lambda_s^{-3}e^{-\beta v_s^\text{ext}(r)}  \frac{\sinh(ip\beta \vert \nabla\psi(r) \vert)}{ip\beta \vert \nabla\psi(r)\vert} \bigg).
\end{align}
Using the operator inverse of the Coulomb potential,
\begin{align}
V^{-1}(r,r')=-\nabla\big(\epsilon_{0}\nabla\delta(r-r')\big),
\end{align}
and assuming that $\nabla\psi$ vanishes at the boundary of the integration domain, the grand canonical partition function can be written as a functional integral over the HS field, 
\begin{align}\label{GrandCanonicalPartitionFunction}
\mathcal{Z}_{\text{gc}}(\mu_{j})=\int D\psi e^{-\beta S[\psi]},
\end{align}
weighted by the exponential of the non-linear field-action, 
\begin{align}\label{action}
S[\psi]=\int_{r}\Bigg(&\frac{\epsilon_0}{2}(\nabla\psi(r))^{2} 
-\sum_{i=a/c}\beta^{-1}\lambda_{i}\Lambda_{i}^{-3}e^{-\beta v_{i}^\text{ext}(r)}e^{-iq_{i}\beta\psi(r)}\nonumber\\
&-\beta^{-1}\lambda_{s}\Lambda_{s}^{-3}e^{-\beta v_s^\text{ext}(r)}\frac{\sinh(ip\beta \vert \nabla\psi(r)\vert)}{ip\beta \vert \nabla\psi(r) \vert}\Bigg).
\end{align}

Physical information can be extracted from Eq.~\eqref{GrandCanonicalPartitionFunction} by computing functional averages. The expectation value is defined as the functional mean with respect to the above probability distribution,
\begin{align}\label{ExpectationValue}
    \avg{...}\equiv\frac{1}{\zgc}\int D\psi (...)e^{-\beta S[\psi]}.
\end{align}
For instance, the electric potential distribution is obtained from the functional integral Eq.~~\eqref{GrandCanonicalPartitionFunction} by
\begin{align}\label{ElectrostaticPotential}
\phi(r)\equiv i\langle \psi(r)\rangle,
\end{align}
i.e. as the expectation value of the fluctuating field times the imaginary unit. It should be noted that the field $\psi$ is real, as previously stated. However, since the expectation value is calculated with respect to a complex probability distribution, as defined in Eq.~\eqref{ExpectationValue}, the average value, $\avg{\psi(r)}$, is imaginary. The variable $\phi$ is the electric potential, since it is the field configuration that exactly solves the Poisson equation for the ensemble-averaged particle density,\cite{netzPoissonBoltzmannFluctuationEffects2000}
\begin{align}\label{Defn:ElectrostaticPotential}
    -\nabla\left(\epsilon(r)\nabla\phi\right)=\sum_{i=a/c}q_i\avg{n_i(r)},
\end{align}
where $\epsilon(r)$ is the permittivity.

\subsection{Transformation to a variational functional}\label{SubSec:TransitionVariational}
The functional Eq.~\eqref{GrandCanonicalPartitionFunction} is still impossible to solve exactly. However, we are not interested in the value of the partition function. We are interested in the value of physical quantities that can be computed from Eq.~\eqref{GrandCanonicalPartitionFunction} utilizing functional averages. In this section, we explain the necessary steps to derive a variational functional from Eq.~\eqref{GrandCanonicalPartitionFunction}, which, when minimized, yields the correct equilibrium distributions. We want to derive a variational functional for electrostatic potential and electrolyte densities. Note that the derivation of the density functional below is similar to the derivation of density functionals in the field of classical fluids,\cite{evansNatureLiquidvapourInterface1979} and follows the approach originally developed for quantum mechanical systems.\cite{hohenbergInhomogeneousElectronGas1964,merminThermalPropertiesInhomogeneous1965}

From now on the explicit dependencies on $\mu_j$ are omitted. The functional mean of the electrostatic potential, Eq.~\eqref{ElectrostaticPotential}, can be computed from Eq.~\eqref{GrandCanonicalPartitionFunction} by introducing an auxiliary charge density $\rho_{\text{aux}}$, which couples linearly to $\psi$, so that Eq.~\eqref{GrandCanonicalPartitionFunction} becomes a functional of $\rho_{\text{aux}}$,
\begin{align}\label{GrandCanonicalPartitionSum}
\tilde{\mathcal{Z}}_{\text{gc}}[\rho_\text{aux}]=\int D\psi e^{-\beta S[\psi]+i \beta\int_r \rho_\text{aux}(r) \psi(r)} \ .
\end{align}
The functional $\tilde{\Omega}_\text{sol}=\beta^{-1}\log\tilde{\zgc}$ has more degrees of freedom than Eq.~\eqref{GrandCanonicalPartitionFunction}, and is equal to it only when $\rho_{\text{aux}}=0$. In this article, the round brackets are used to denote the argument of functions, while the square brackets are used to denote the argument of functionals. Taking the functional derivative with respect to $\rho_\text{aux}$ allows calculating the electric potential, as defined in Eq.~\eqref{Defn:ElectrostaticPotential}, for a given $\rho_\text{aux}$,
\begin{align}\label{DefinitionElectrostaticPotential}
\phi(r)&= -\frac{\delta \tilde{\Omega}_\text{sol}[\rho_\text{aux}]}{\delta \rho_\text{aux}(r)} \nonumber\\
&=\frac{1}{\beta} \frac{1}{\tilde{\zgc}} \frac{\delta\tilde{\zgc}}{\delta \rho_\text{aux}(r)} = i\langle \psi(r) \rangle.
\end{align}
Note that the physical equilibrium potential $\phi^\text{eq}$ is obtained only for $\rho_\text{aux }=0$, i.e.,
\begin{align}\label{PhysicalElectrostaticPotential}
    \phi^\text{eq}(r)= -\at{\frac{\delta \tilde{\Omega}_\text{sol}[\rho_\text{aux}]}{\delta \rho_\text{aux}(r)}}{\rho_\text{aux }=0}.
\end{align}
In a similar way, the average electrolyte density can be computed by linearly coupling an auxiliary potential $v_j^\text{aux}(r)$ to the particle density of type $j$. Physically, $v_j^\text{aux}(r)$ simply represents an additional external potential, which can be interpreted  as a variation of the actual $v_j^\text{ext}(r)$,
\begin{align}
    v_j^\text{ext}(r)\to v_j^\text{ext}(r) + v_j^\text{aux}(r).
\end{align}
The action Eq.~\eqref{action} including $v^\text{aux}_j$ has the form
\begin{align}\label{ActionInAuxillaryField}
S[\psi]=\int_{r}\Bigg(&\frac{\epsilon_0}{2}(\nabla\psi(r))^{2} \nonumber \\
&-\sum_{i=a/c}\beta^{-1}\Lambda_{i}^{-3}e^{\beta(\mu_i-v_i^\text{ext}(r)-v_i^\text{aux}(r))}e^{-iq_{i}\beta\psi(r)}\nonumber\\
&-\beta^{-1}\Lambda_{s}^{-3}e^{\beta(\mu_s-v_s^\text{ext}(r)-v_s^\text{aux}(r))}\frac{\sinh(ip\beta \vert \nabla\psi(r)\vert)}{ip\beta \vert \nabla\psi(r) \vert}\Bigg).
\end{align}
Thus, we have converted the partition function (Eq.~\eqref{Defn:PartitionFunction}) into a functional of four auxiliary fields, i.e. three $v_j^\text{aux}(r)$ and one $\rho_\text{aux}(r)$. From this, we obtain the grand potential \emph{functional}, $\tilde{\Omega}_\text{sol}[v_j^\text{aux}(r)\rho_\text{aux}(r)]$. The ensemble averaged particle densities, for a given $v_j^\text{ext}(r)$ and $v_j^\text{aux}(r)$, can be computed by functional derivation of the grand potential functional,
\begin{align}\label{ComputingDensityFromPartitionFunction}
    n_{j}(r)&= \frac{\delta\tilde{\Omega}_\text{sol}[v_j^\text{aux}(r),\rho_\text{aux}(r)]}{\delta v_j^\text{aux}(r)}\\
    &=-\frac{1}{\beta}\frac{1}{\tilde{\zgc}}\frac{\delta\tilde{\zgc}[v_j^\text{aux}(r)(r),\rho_\text{aux}(r)]}{\delta v_j^\text{aux}(r)},
\end{align}
where the equilibrium densities are obtained by setting $v_j^\text{aux}(r)=0$, i.e.
\begin{align}\label{EquilibriumDensity}
    n^\text{eq}_{j}(r)= \at{\frac{\delta\tilde{\Omega}_\text{sol}[v_j^\text{aux}(r),\rho_\text{aux}(r)]}{\delta v_j^\text{aux}(r)}}{v_j^\text{aux}(r)=0}.
\end{align}
In total, this yields four conjugated pairs,
\begin{align}\label{ConjugatedPairs}
    \rho_\text{aux}(r) \longleftrightarrow \phi(r), \quad v^\text{aux}_j(r) \longleftrightarrow n_j(r) \quad \text{with } j\in a,c,s,
\end{align}
of one physical quantity ($\phi,n_j$) and one auxiliary field ($\rho_\text{aux},v_j^\text{aux}$) that are related through a functional derivative of the grand potential functional.

We aim to develop a variational functional that will yield the correct thermodynamic equilibrium density and potential distributions by solving the Euler-Lagrange equations. This means finding the field configuration for which the functional derivative of the variational functional is zero. To obtain a variational principle for $\phi(r)$ and $n_j(r)$, we use two Legendre transformations that substitute the dependence of $\tilde{\Omega}_\text{sol}[v_j^\text{aux}(r),\rho_\text{aux}(r)]$ on $v_j^\text{aux}(r)$ and $\rho_\text{aux}(r)$ by their conjugates, as in Eq.~\eqref{ConjugatedPairs}. The variational solution functional is given by:
\begin{align}\label{GrandPotentialSolutionFunctional}
\Omega_\text{sol}[n_j(r),\phi(r)]=&\tilde{\Omega}_\text{sol}[v_j^\text{aux}(r),\rho_\text{aux}(r)] \nonumber \\+&\int_{r}\rho_\text{aux}(r)\phi(r) -\sum_{j=a/c/s}\int_{r}v^\text{aux}_j(r)n_{j}(r),
\end{align}
where $\phi$ and $\rho_\text{aux}$ are related by Eq.~\eqref{DefinitionElectrostaticPotential} and $n_j(r)$ and $v_j^\text{aux}$ by Eq.~\eqref{ComputingDensityFromPartitionFunction}. Recognizing that $\rho_\text{aux}$ and $v_j^\text{aux}$ are functions of $\phi$, the functional Eq.~\eqref{GrandPotentialSolutionFunctional}  satisfies
 \begin{align}
	\frac{\delta \Omega_\text{sol}[v_j^\text{aux}(r),\phi(r)]}{\delta \phi(r)}=\rho_\text{aux}(r).
\end{align}
Since the physical electrostatic potential $\phi^\text{eq}(r)$ (Eq.~\eqref{PhysicalElectrostaticPotential}) corresponds to $\rho_\text{aux}=0$, we obtain
\begin{align}\label{EquationOfState}
\frac{\delta \Omega_\text{sol}[v_j^\text{aux},\phi^\text{eq}(r)]}{\delta \phi}=0,
\end{align}
 i.e., $\Omega_\text{sol}$ is stationary at the physical (equilibrium) potential. Analogously, Eq.~\eqref{GrandPotentialSolutionFunctional} satisfies
\begin{align}
    \frac{\delta\Omega_\text{sol}[v_j^\text{aux},\phi(r)]}{\delta n_j(r)} = -v_j^\text{aux}(r).
\end{align}
This means the variational derivative of $\Omega_\text{sol}$,
\begin{align}\label{VariationalEqOmegaCDensity}
    \frac{\delta\Omega_\text{sol}[n^\text{eq}_j(r),\phi(r)]}{\delta n_j(r)} =0,
\end{align}
vanishes for $n^\text{eq}_j(r)$, \textit{cf.}  Eq.~\eqref{EquilibriumDensity}, which corresponds to $v_j^\text{aux}(r)=0$.

In summary, from a classical first-principles approach, we have rigorously derived a functional $\Omega_\text{sol}$ that is stationary with respect to $n^\text{eq}_j$ and the $\phi^\text{eq}$,
\begin{align}\label{VariationalEqOmegaCPhi}
    \frac{\delta \Omega_\text{sol}[n_j(r),\phi^\text{eq}(r)]}{\delta\phi(r)} = 0, \quad \frac{\delta \Omega_\text{sol}[n^\text{eq}_j(r),\phi(r)]}{\delta n_j(r)} = 0.
\end{align}
The variational equations given in Eq.~\eqref{VariationalEqOmegaCPhi} result in differential equations known as the Euler-Lagrange equations for \(\phi\) and \(n_j\). We denote the variational functional $\Omega_\text{sol}[n_j(r),\phi^\text{eq}(r)]$ by the same symbol as the grand-potential function (Eq.~\eqref{GrandPotentialFunction}) since evaluating the functional at the physical equilibrium configurations, $\phi^\text{eq}$ and $n^\text{eq}_j$, yields the value of the grand potential function at given $\mu_j$,
\begin{align}
    \Omega_\text{sol}[n^\text{eq}_j(r),\phi^\text{eq}(r)] = \Omega_\text{sol}(\mu_j),
\end{align}
where the $\mu_j$ dependence in the l.h.s. is not explicitly stated.
\subsection{Derivation of the 1L-LDA functional}\label{SubSec:OneLoopAndLocalDensity}
Thus far, no approximations have been used in the formalism. This subsection explains the explicit calculation of $\Omega_\text{sol}$ based on two approximations. We split the two Legendre transformations in Eq.~\eqref{GrandPotentialSolutionFunctional} into two steps: in Sec.~\ref{SubSub:OneLoopApproximation}, we introduce the one loop (1L) approximation to obtain an explicit form of the Legendre transformation for the electrostatic potential. In Sec.~\ref{SubSub:LDA}, we perform the second Legendre transformation for the electrolyte densities. Furthermore, a local-density approximation (LDA) is applied to obtain an analytical expression for the correlation functional. This allows finding a local free energy functional including coulombic correlation effects for electrolyte solutions in the presence of an external potential.

\subsubsection{The One-Loop Approximation}\label{SubSub:OneLoopApproximation}
The first Legendre transformation, Eq.~\eqref{GrandPotentialSolutionFunctional}, results in a variational functional soley for the electric potential,
\begin{align}\label{Generating Functional}
\Gamma[v_j^\text{aux}(r),\phi(r)]=\tilde{\Omega}_\text{sol}[v_j^\text{aux}(r),\rho_\text{aux}(r)]+\int_{r}\phi(r)\rho_\text{aux}(r).
\end{align}
To compute $\Gamma$, we insert $\Omega_\text{sol}$ from Eq.~(\ref{GrandPotentialFunction}), including the auxiliary fields, into Eq.~(\ref{Generating Functional}) and exponentiate both sides, leading to
\begin{align}
&e^{-\beta \Gamma[v_j^\text{aux},\phi(r)]}=\int D\psi \nonumber\\
&\exp\left(-\beta S[\psi]+\beta \int_r \rho_\text{aux}(r) \big(i\psi(r)-\phi(r)\big)\right).
\end{align} 
Shifting the integral over all field configurations, $\psi$, to one over all fluctuations, $\delta\psi$, around the mean, $\langle\psi(r)\rangle$, 
\begin{align}\label{AuxillaryDecomposition}
\psi(r)=\langle\psi(r)\rangle+\delta\psi(r),
\end{align}
yields 
\begin{align}\label{EffectiveAction}
&e^{-\beta\Gamma[v_j^\text{aux},\phi(r)]}=\int D\delta\psi \nonumber\\
&\exp\bigg(-\beta S[i^{-1}\phi+\delta\psi]+i\beta\int_{r}\delta\psi(r)\rho_{\text{aux}}(r)\bigg).
\end{align}
There are several different ways to approximate $\Gamma$ at this point. When fluctuations $\delta \psi$ are entirely neglected, implying that electrolyte correlations are disregarded
\begin{align}
\Gamma[v_j^\text{aux},\phi(r)]=S[i^{-1}\phi(r)].
\end{align}
Theories that neglect correlations are referred to as mean-field (MF) theories, in which particle-particle interactions are replaced by interactions with an average potential.\cite{abrashkinDipolarPoissonBoltzmannEquation2007}

If the action \( S \) is large, the functional integral, described by Eq.~\eqref{EffectiveAction}, can be approximated using the saddle-point approximation, where the action is expanded around the saddle-point configuration that satisfies

\begin{align}\label{Saddlepoint}
\frac{\delta S}{\delta\psi(r)}-i\;\rho_{\text{aux}}(r) =0.
\end{align}
This method relies on an expansion parameter, \(\upsilon\), which is small and serves as a pre-factor of the action. Following Netz et al.,\cite{netzElectrostatisticsCounterionsPlanar2001} the action is multiplied  by \(1/\upsilon\), which acts as our expansion parameter and helps distinguish terms of different orders and is set to one in the end. In Section \ref{SubSec:Applicability}, we demonstrate that for the 1L-LDA functional, \(\upsilon\) is inversely proportional to the Debye length. Therefore, the expansion becomes exact in the limit of infinite dilution, and the accuracy is determined by the highest density in the system. 

For the expansion, it is convenient to redefine $\delta\psi\to\upsilon^{-1/2}\delta\psi$ in Eq.~\eqref{EffectiveAction}. The expansion of $S$ around $i^{-1}\phi$ then reads
\begin{align}\label{ExpansionAroundMeanField}
    &e^{-\beta\Gamma[v_j^\text{aux},\phi(r)]}=\int D\delta\psi \exp\bigg(-\upsilon^{-1}\beta S[i^{-1}\phi] \nonumber \\
&-\upsilon^{-1/2}\beta\int_{r}\left.[\frac{\delta S}{\delta\psi(r)}-i\;\rho_\text{aux}(r)]\right \vert_{\psi=i^{-1}\phi}\delta\psi(r)\nonumber\\
&-\frac{\beta}{2}\int_{r,r'}\delta\psi(r)\left.\frac{\delta^{2}S}{\delta\psi(r)\delta\psi(r')}\right\vert_{\psi=i^{-1}\phi}\delta\psi(r')+ \mathcal{O}(\upsilon^{1/2})\bigg).
\end{align}
The first term in Eq.~\eqref{ExpansionAroundMeanField} is the action evaluated at the expectation value of the HS field $\avg{\psi(r)}=i^{-1}\phi(r)$ and the second term is the the saddle point equation, which vanishes when evaluated at the expansion point \footnote{The expansion point $\phi$ is not exactly equal to the saddle point configuration of $S$. However, it can be shown that to leading order in $\upsilon$ they are equal.\cite{netzPoissonBoltzmannFluctuationEffects2000}}. The second order expansion in the third term is called the 1L expansion. This designation comes from the fact that,within a diagrammatic expansion of the action, the second order term contains exactly all diagrams with one loop.\cite{zinn-justinQuantumFieldTheory2021}  

Plugging Eq.~\eqref{Saddlepoint} into Eq.~\eqref{ExpansionAroundMeanField} and neglecting all terms of order $\mathcal{O}(\upsilon^{1/2})$ and higher, we arrive at
\begin{align}\label{DerivationOneLoop}
&e^{-\beta\Gamma[v_j^\text{aux},\phi(r)]}\approx e^{-\upsilon^{-1}\beta S[i^{-1}\phi]}\int D\delta\psi \nonumber \\
&\exp\left(-\frac{\beta}{2}\int_{r,r'}\delta\psi(r)\left.\frac{\delta^{2}S}{\delta\psi(r)\delta\psi(r')}\right\vert_{\psi=i^{-1}\phi}\delta\psi(r')\right) \nonumber\\
&=e^{-\upsilon^{-1}\beta S[i^{-1}\phi]}\left(\det\beta\left.\frac{\delta^{2}S}{\delta\psi(r)\delta\psi(r')}\right\vert_{\psi=i^{-1}\phi}\right)^{-1/2}\nonumber\\
&=\exp\Biggl(-\upsilon^{-1}\beta S[i^{-1}\phi]-\frac{1}{2}\text{tr}\log\beta\left.\frac{\delta^{2}S}{\delta\psi(r)\delta\psi(r')}\right\vert_{\psi=i^{-1}\phi}\Biggr).
\end{align}
Here, a general expression for a Gaussian functional integral in terms of the determinant of the kernel was used,\cite{netzPoissonBoltzmannFluctuationEffects2000} neglecting an irrelevant constant factor. By re-exponentiating the determinant and using the matrix identity $\log(\det(A))=\text{tr}(\log(A))$, one arrives at the final expression. A comparison of the left and right sides of Eq.~(\ref{DerivationOneLoop}), gives the result for $\Gamma$ in the 1L approximation,
\begin{align}\label{EffectiveActionInOneLoop}
\Gamma[v_j^\text{aux},\phi(r)]=S[i^{-1}\phi]+\frac{\upsilon}{2\beta}\text{tr}\log\beta G^{-1}[i^{-1}\phi],
\end{align}
with
\begin{align}\label{DefnGreenFunction}
    G^{-1}(r,r')=\left.\frac{\delta^{2}S}{\delta\psi(r)\delta\psi(r')}\right\vert_{\psi=i^{-1}\phi}.
\end{align}
The first term on the r.h.s. in Eq.~\eqref{EffectiveActionInOneLoop},  evaluated at $i^{-1}\phi$ gives
\begin{align}
S[i^{-1}\phi]&=\int_{r}-\frac{\epsilon_0}{2}(\nabla\phi)^{2}\nonumber\\
&-\sum_{i=a/c}\beta^{-1}e^{\beta (\mu_i-v_i^\text{ext}(r)-v_i^\text{aux}(r))}\Lambda_{i}^{-3}e^{-q_{i}\beta\phi(r)}\nonumber\\
&-\beta^{-1}e^{\beta (\mu_s-v_s^\text{ext}(r)-v_s^\text{aux}(r))}\Lambda_{s}^{-3}\frac{\sinh(p\beta\abs{\nabla\phi(r)})}{p\beta\abs{\nabla\phi(r)}}.
\end{align}
The second term in Eq.\eqref{EffectiveActionInOneLoop}, contains Eq.\eqref{DefnGreenFunction} , which is computed in appendix \ref{Derivation of the second variation} and yields
\begin{align}\label{InverseGreen}
    G^{-1}(r,r') =& -\nabla_{r'}\left(\epsilon_G(r')\nabla_{r'}\delta(r-r')\right) \nonumber\\
    &+\sum_{i=a/c}q_{i}^{2}\beta\lambda_{i}\Lambda_{i}^{-3}e^{-q_{i}\beta\phi(r')}\delta(r-r'),
\end{align}
with
\begin{align}
    \epsilon_G(r)=\epsilon_0+\lambda_{s}\Lambda_{s}^{-3}p^{2}\beta\frac{\sinh(p\beta \vert \nabla \phi \vert)}{p\beta \vert \nabla \phi \vert}(\mathcal{L}^{2}+\mathcal{L}'),
\end{align}
where, omitting the argument, $\mathcal{L}\equiv\mathcal{L}(u)=\coth(u)-1/u$ and $\mathcal{L}'\equiv\mathcal{L}'(u)$, are the the Langevin function and its derivative, respectively and $u=p\beta \vert \nabla \phi \vert$. Throughout the paper it is not necessary to determine the value of the $\text{tr}\log$ expression in Eq.~\eqref{EffectiveActionInOneLoop}, which depends on $G^{-1}$. In any instance, knowledge of the Green's function will suffice to compute all quantities, \textit{cf.} Eqs.~\eqref{ionicScalingLength} and \eqref{solventScalingLength}. If one insert Eq.\eqref{InverseGreen} into the definition for the Green's function,
\begin{align}
\int_{r''}G^{-1}(r,r'')G(r'',r')=\delta(r-r'),
\end{align}
one gets
\begin{align}\label{GreenFunction}
    &-\nabla_{r'}\left(\epsilon_G(r')\nabla_{r'}G(r,r')\right)\nonumber\\
    &+\sum_{i=a/c}q_{i}^{2}\beta\lambda_{i}\Lambda_{i}^{-3}e^{-q_{i}\beta\phi(r')}G(r,r') = \delta(r-r').
\end{align}
Note that for $p=0$, which means that the solvent is apolar, $G^{-1}$ reduces to the differential operator whose solution is a screened Coulomb potential. Thus, the Green's function entails the screening. The Green's function describes the electrostatic correlations in the system and is therefore also referred to as the correlation function.

We now perform the second Legendre transformation in Eq.~\eqref{GrandPotentialSolutionFunctional} to obtain a variational functional of both the electrostatic potential and the electrolyte densities. To this end, the expression for the densities, Eq.~\eqref{ComputingDensityFromPartitionFunction}, must be inverted to a function of $v_j^\text{aux}$ and substituted into
\begin{align}
    \Omega_\text{sol}[n_j,\phi]=\Gamma[v_j^\text{aux},\phi] - \sum_{j={a,c,s}} \int_r v_j^\text{aux}(r)n_j(r).
\end{align}
For details regarding the derivation of $v_j^\text{aux}$, we refer the reader to appendix \ref{DerivationChemicalPotentials}. The results for $v_j^\text{aux}$ as a function of $n_j$ are
\begin{align}
	v_i^\text{aux}(r)=&\mu_{i}-\mu_i^\text{ref}-\beta^{-1}\log\left(\frac{n_{i}(r)/n_i^\text{ref}}{l_{i}(r)/l_i^\text{ref}}\right) \nonumber \\
 &-q_{i}\phi(r)-v_i^\text{ext}(r), \label{ionChemicalPotential}\\
	v_s^\text{aux}(r)=&\mu_{s}-\mu_s^\text{ref}-\beta^{-1}\log\left(\frac{n_{s}(r)/n_s^\text{ref}}{l_{s}(r)/l_s^\text{ref}}\right) \nonumber\\
& +\beta^{-1}\log\left(\frac{\sinh(p\beta\vert\nabla\phi(r)\vert)}{p\beta\vert\nabla\phi(r)\vert}\right)-v_s^\text{ext}(r).\label{solventChemicalPotential}
\end{align}
For practical calculations, we introduced reference states with corresponding reference chemical potentials, denoted as $\mu_j^\text{ref}$, where the electrostatic potential is zero and the density is constant and equal to an arbitrarily chosen reference value. 
In the third terms on the right hand sides of Eqs.~(\ref{ionChemicalPotential}) and (\ref{solventChemicalPotential}), densities are rescaled by dimensionless parameters, which are given as follows,
\begin{align}
l_{i}(r)&\equiv1-\upsilon\frac{\beta q_{i}^{2}}{2}G(r,r),\label{ionicScalingLength}\\
l_{s}(r)&\equiv1+\upsilon\frac{\beta p^{2}}{2}(\mathcal{L}^{2}+\mathcal{L}')\nabla^{2}G(r,r).\label{solventScalingLength}
\end{align}
These parameters encapsulate the corrections at the level of the 1L expansion and depend primarily on the equal-point correlation function. Thus, the introduction of $l_{i}$ and $l_{s}$ facilitates the interpretation of correlation effects using a single local parameter for each particle species depending on the equal-point correlation function $G(r,r)$ for ions and the Laplacian of the equal-point correlation function $\nabla^{2}G(r,r)$ for the solvent. The way how the correlation parameters $l_j/l^\text{ref}_j$ enter Eq.~\eqref{ionChemicalPotential} and \eqref{solventChemicalPotential}, shows that $l_j/l^\text{ref}_j$ can be physically interpreted as the inverse of activity coefficients, \textit{cf.} Eqs.~\eqref{ionChemicalPotential} and \eqref{solventChemicalPotential}, as discussed in more detail in Sec.~\ref{SubSubSec:CorrelationPotential}. Finally, the fourth terms in Eq.~(\ref{ionChemicalPotential}) and Eq.~(\ref{solventChemicalPotential}) include the conventional MF coupling to the electric potential.\cite{huangHybridDensitypotentialFunctional2021}

The expressions for $v_j^\text{aux}$, given in Eqs.~\eqref{ionChemicalPotential} and \eqref{solventChemicalPotential}, inserted into the Legendre transformation Eq.~\eqref{GrandPotentialSolutionFunctional}, yield
\begin{align}
    \Omega_\text{sol}[n_j(r),\phi(r)]=&\mathcal{F}_\text{sol}[n_j(r),\phi(r)]- \sum_{j=a/c/s}\mu_j\int_r n_j(r) ,
\end{align}
where $\mathcal{F}_\text{sol}$ is the free-energy solution functional of $n_j$ and $\phi$, including the external potential $v_j^\text{ext}$,
\begin{align}\label{ResultVariationalFunctional}
    \mathcal{F}_\text{sol}[n_j(r),\phi(r)]=\mathcal{F}_\text{sol}^\text{id}+\mathcal{F}^\text{mf}_\text{sol}+\mathcal{F}^\text{corr,1L}_\text{sol} + \sum_j\int_r v_j^\text{ext}(r)n_j(r).
\end{align}
The free energy functional of Eq.~\eqref{ResultVariationalFunctional} splits into universal functionals of $n_j$ and $\phi$, but independent of $v_j^\text{ext}$. This is no surprising result, since the existence of a universal functional, independent of any external potential is a central result from classical but also quantum DFT.\cite{hansenTheorySimpleLiquids2006,kohnSelfConsistentEquationsIncluding1965} Here, $\mathcal{F}^\text{id}_\text{sol}$ is the ideal gas functional,
\begin{align}\label{idealGasFreeEnergy}
    \mathcal{F}^\text{id}_\text{sol} =& \int_r \sum_j \beta^{-1} n_j(r) \left( \log \left(n_j(r)/n_j^\text{ref}\right) -1 \right)+\int_r \sum_j \mu^\text{ref}_j n_j(r), 
\end{align}
Furthermore we find an excess part that describes interactions between particles. The latter is further split into the usual excess MF functional 
\begin{align}\label{MeanFieldExcessFreeEnergy}
    \mathcal{F}^\text{mf}_\text{sol}=&\int_r \Bigg( -\frac{\epsilon_0}{2}(\nabla\phi)^2 + \sum_i n_i(r)q_i  \phi(r) \nonumber\\
    &-n_s(r)\beta^{-1} \log \left( \frac{\sinh(p\beta\vert\nabla\phi\vert)}{p\beta\vert\nabla\phi\vert} \right)\Bigg)
\end{align}
and a novel 1L correlation functional
\begin{align}\label{CorrelationFunctional}
    \mathcal{F}^\text{corr,$1$L}_\text{sol} =\int_r\sum_j n_j(r)\epsilon_{\text{corr},j}(r)  +\frac{\upsilon}{2\beta}\text{tr}\log\beta G^{-1}[\phi(r)] \ .
\end{align}
The first term on the r.h.s. of Eq.~\eqref{CorrelationFunctional} is a functional contribution of the electrolyte densities of the form
\begin{align}
    \epsilon_{\text{corr},j}(r)=-\beta^{-1}\left(\log\left(l_j(r)/l_j^\text{ref}\right)+\frac{1}{l_{j}(r)}-1\right).
\end{align}
The second term on the r.h.s. of Eq.~\eqref{CorrelationFunctional} is a functional contribution of the electric potential. 

The expressions for $l_j$, Eqs.~(\ref{ionicScalingLength}) and (\ref{solventScalingLength}), reveal that the $l_j$ approach a value of one when correlation effects are small. In this limit $\mathcal{F}_\text{sol}$ reduces to the MF expression\cite{huangGrandCanonicalModelElectrochemical2021} as expected.

The differential equation for the Green's function, Eq.~\eqref{GreenFunction},  written with electrolyte densities, reads
\begin{align}\label{GreenFunctionCanonical}
    &-\nabla_{r'}\left(\epsilon_G(r')\nabla_{r'}G(r,r')\right)\nonumber\\
    &+\sum_{i=a/c}q_{i}^{2}\beta n_i(r)G(r,r') = \delta(r-r') \ ,
\end{align}
with a permittivity for $G$,
\begin{align}\label{PermittivityForCorrelations}
    \epsilon_G(r)=\epsilon_0+p^{2}\beta n_s(r)(\mathcal{L}^{2}+\mathcal{L}') \ .
\end{align}
Note that, so far, the only approximation employed was the 1L expansion. The free energy functional in Eq.~\eqref{ResultVariationalFunctional} is exact up to $\mathcal{O}(\upsilon)$. 

\subsubsection{The Local-Density-Approximation}\label{SubSub:LDA}

In the presence of arbitrary external potentials $v_\text{ext}$, where densities are non-uniform, the equal point correlation function, given in equations \eqref{ionicScalingLength} and \eqref{solventScalingLength}, cannot be determined analytically as a function of local densities. For instance, at a metal--electrolyte interface, the fields are spatially dependent and the solution to Eq.~\eqref{GreenFunctionCanonical} must be obtained self-consistently in conjunction with the variational equations~\eqref{VariationalEqOmegaCPhi}. 

We opt for a local-density approximation (LDA), where we derive $G$ and hence $l_j$ assuming constant fields. The Green's function is then only a function of the distance $G(r,r')=G(r-r')$ and can be obtained analytically.  Suppressing the spatial dependence of the densities in Eq.~\eqref{GreenFunctionCanonical}, the differential equation reads

\begin{align}\label{DiffEqGreen}
\big(-\nabla_{r'}^{2}+\lambda_D^{-2}\big)G(r-r')=\frac{1}{\epsilon_G}\delta(r-r'),
\end{align}
which is solved by
\begin{align}
    G(r-r')=\frac{1}{4\pi\epsilon_G}\frac{e^{\abs{r-r'}/\lambda_D}}{\abs{r-r'}},
\end{align}
where
\begin{align}\label{DebyeLength}
\lambda_D = \sqrt{\frac{\epsilon_G}{\beta \sum_{i=a/c}q_{i}^{2}n_{i}}},
\end{align}
is the Debye length for $G$ containing the correlation permittivity $\epsilon_G$, given in Eq.~\eqref{PermittivityForCorrelations}.

Using the LDA, we avoid the self-consistent solution of Eq.~\eqref{GreenFunctionCanonical} in an external potential but only account for bulk electrolyte correlation effects. This simplification is similar to the assumption in electronic DFT where the local density functional is derived from the homogeneous electron gas.\cite{kohnSelfConsistentEquationsIncluding1965,merminThermalPropertiesInhomogeneous1965,hohenbergInhomogeneousElectronGas1964}

The correlation parameters  Eqs.~\eqref{ionicScalingLength} and  \eqref{solventScalingLength} , inside of the correlation functional in Eq.~\eqref{CorrelationFunctional}, depend on the equal-point Green's function, which diverges at small distances, formally known as ultraviolet divergence.\cite{zinn-justinQuantumFieldTheory2021} This is a typical problem of point-charge models, arising from the fact that the electrical energy of two charged particles diverges when they can come arbitrarily close to each other. 

For this purpose, we introduce two short distance cutoffs in the computation of the equal point correlation function in Eqs.~\eqref{DerivationGreenFunction} and \eqref{DerivationGreenFunction2}: $a_i$ for the computation of $l_i$ in Eq.~\eqref{ionicScalingLength} and $a_s$ for the computation of $l_s$ in Eq.~\eqref{solventScalingLength}. It will become clear, in Sec.~\ref{Sec: Results}, that $a_i$ is related to coulombic screening, while $a_s$ is related to the dielectric properties of the electrolyte. Furthermore, it will be discussed how these parameters can be determined from experimental data. Details of the derivation of the correlation parameters in LDA are presented in Appendix \ref{Appendix:Local-density}. We obtain
\begin{align}
l_{i}(r)&=1-\upsilon\frac{\beta q_{i}^{2}}{4\pi^{2}\epsilon_G(r)}\times \nonumber\\
&\times\left(\frac{2\pi}{a_i}-\frac{1}{\lambda_D(r)}\arctan\left(2\pi\frac{\lambda_D(r)}{a_i}\right)\right), \label{ScalingParameterIon}\\
l_{s}(r)&=1-\upsilon\frac{\beta p^{2}}{4\pi^{2}\epsilon_G(r)}(\mathcal{L}^{2}+\mathcal{L}')\times
\nonumber\\
&\times\left(\frac{8\pi^{3}}{3a_s^{3}}-\frac{2\pi}{\lambda_D(r)^{2}a_s}+\frac{1}{\lambda_D(r)^{3}}\arctan\left(2\pi\frac{\lambda_D(r)}{a_s}\right)\right)\label{ScalingParameterSolvent},
\end{align}
where we restored the spatial dependence of $\epsilon_G(r)$ and $\lambda_D(r)$ given in Eqs.~\eqref{PermittivityForCorrelations} and \eqref{DebyeLength}. 
The correlation functional described in Eq.~\eqref{CorrelationFunctional}, combined with the local density approximation used in Eqs.~\eqref{ScalingParameterIon} and \eqref{ScalingParameterSolvent}, is referred to as the 1L-LDA functional, denoted as \(\mathcal{F}^\text{corr,1L-LDA}_\text{sol}\).

\subsection{Functional Derivatives of the 1L-LDA functional}\label{SubSec:FunctionalDerivative}
For practical purposes, functional derivatives of the LDA correlation functional are required. The correlation functional alters the Euler-Lagrange equation for densities by introducing an additional correlation potential. Simultaneously, the Euler-Lagrange equation for the electrostatic potential is affected by a correlation-induced charge density, which can be integrated into a redefined permittivity.

\subsubsection{The LDA Correlation Charge Density}
The variational equation, Eq.~\eqref{VariationalEqOmegaCPhi}, for  $\phi$ yields a Poisson-type differential equation for the electrostatic potential. According to Eq.~\eqref{ResultVariationalFunctional}, we need to compute three functional derivatives. The ideal gas functional, $\mathcal{F}_\text{sol}^\text{id}$, does not depend on $\phi$,
\begin{align}\label{VariationalEqPhi1}
    \frac{\delta \mathcal{F}_\text{sol}^\text{id}}{\delta\phi(r)} = 0.
\end{align}
The functional derivative of the MF excess functional yields
\begin{align}\label{VariationalEqPhi2}
    \frac{\delta \mathcal{F}^\text{mf}_\text{sol}}{\delta\phi(r)} &= \sum_{i}q_{i}n_{i}(r) \nonumber\\
    &-\nabla\cdot\left(-\epsilon_{0}\nabla\phi(r)-\frac{n_{s}p}{\abs{\nabla\phi}}\mathcal{L}(p\beta\abs{\nabla\phi})\nabla\phi(r)\right).
\end{align}
The functional derivative of the correlation functional is complicated by the dependence of $l_j$ on $\nabla\phi$. Details for the functional derivative are presented in Appendix \ref{App:FunctionalDerivatives}, yielding
\begin{align}\label{LDAVariationalDerivative1}
    \frac{\delta \mathcal{F}^\text{corr,$1$L-LDA}_\text{sol}}{\delta\phi(r)} &= \nonumber \\
    \nabla\cdot\bigg[\frac{n_{s}(r)p}{\abs{\nabla\phi(r)}}&\frac{(\mathcal{L}^{3}+3\mathcal{L}\mathcal{L}'+\mathcal{L}'')}{\mathcal{L}^{2}+\mathcal{L}'}(1-l_{s}(r))\nabla\phi(x)\bigg].
\end{align}
Inserting Eqs.~\eqref{VariationalEqPhi1}, \eqref{VariationalEqPhi2} and \eqref{LDAVariationalDerivative1} into Eq.~\eqref{VariationalEqOmegaCPhi}  yields a Poisson equation, which can be written such that the 1L-LDA functional contributes with an effective correlation charge density on the r.h.s,
\begin{align}\label{CorrelationAsChargeDensity}
    &-\nabla\left(\epsilon^\text{MF}(r)\nabla\phi(r)\right)= \sum_{i}q_{i}n_{i}(r)  \nonumber\\
    &\quad +\nabla\cdot\bigg[\frac{n_{s}(r)p}{\abs{\nabla\phi(r)}}\frac{(\mathcal{L}^{3}+3\mathcal{L}\mathcal{L}'+\mathcal{L}'')}{\mathcal{L}^{2}+\mathcal{L}'}(1-l_{s}(r))\nabla\phi(x) \bigg] \ ,
\end{align}
with MF permittivity
\begin{align}\label{MFDielectricConstant}
   \epsilon^\text{MF}(r) =\epsilon_0+\frac{n_{s}p}{\abs{\nabla\phi}}\mathcal{L}(p\beta\abs{\nabla\phi}) \ .
\end{align}
Alternatively, the correlation-induced charge density can be incorporated into the permittivity so that the Poisson equation takes the simple form
\begin{align}\label{PoissonEq1L}
    -\nabla\left(\epsilon^\text{1L}(r)\nabla\phi(r)\right) = \sum_{i}q_{i}n_{i}(r) \ ,
\end{align}
with the 1L permittivity
\begin{align}\label{1LDielectricConstant}
    \epsilon^\text{1L}(r) = \epsilon_0+\frac{n_{s}p}{\abs{\nabla\phi}}\left( \mathcal{L} +\frac{(\mathcal{L}^{3}+3\mathcal{L}\mathcal{L}'+\mathcal{L}'')}{\mathcal{L}^{2}+\mathcal{L}'}(1-l_{s})\right).
\end{align}

\subsubsection{The LDA Correlation Potential}\label{SubSubSec:CorrelationPotential}
The Euler-Lagrange equations for electrolyte ion and solvent densities, cf. Eq.~\eqref{VariationalEqOmegaCPhi}, yield Boltzmann-like relations. Functional differentiation of $\mathcal{F}_\text{sol}^\text{id}$ with respect to the densities gives
\begin{align}\label{VariationalEq1}
    \frac{\delta \mathcal{F}_\text{sol}^\text{id}}{\delta n_j(r)} = \beta^{-1}\log\left(n_j(r)/n_j^\text{ref}\right)+\mu_j^\text{ref}.
\end{align}
The functional derivatives of the MF excess functional with respect to the densities depend on the type of particles since ions interact differently with the electrostatic potential than solvent molecules. For ions, the functional derivative yields
\begin{align}\label{VariationalEq2}
    \frac{\delta \mathcal{F}^\text{mf}_\text{sol}}{\delta n_i(r)} = q_i \phi(r)\ ,
\end{align}
whereas the derivative for the solvent density yields
\begin{align}\label{VariationalEq3}
    \frac{\delta \mathcal{F}^\text{mf}_\text{sol}}{\delta n_s(r)} = -\beta^{-1} \log \left( \frac{\sinh(p\beta\vert\nabla\phi\vert)}{p\beta\vert\nabla\phi\vert} \right).
\end{align}
The functional derivative of the LDA correlation functional is more difficult to calculate due to the dependence of $l_j$ on $n_j$, which is presented in appendix \ref{App:FunctionalDerivatives}. Neglecting higher orders than $\mathcal{\upsilon}$, the functional derivative gives\begin{align}\label{LDAVariationalDerivative2}
    \frac{\delta \mathcal{F}^\text{corr,$1$L-LDA}_\text{sol}}{\delta n_j(r)} \approx -e_0 \phi_j^\text{corr}(r) 
\end{align}
with a correlation potential,
\begin{align}
    \phi_j^\text{corr}(r) = (\beta e_0)^{-1} \log \left(l_j(r)/l^\text{ref}_j\right),
\end{align}
for both ion and solvent species. The overall Euler-Lagrange equation for the ion densities, Eq.~\eqref{VariationalEqOmegaCPhi}, resolving for $n_j$, yields Boltzmann-like relations,
\begin{align}
    n_i(r) & = n_i^\text{ref} \exp{\left(\beta \left((\mu_i-\mu_i^\text{ref}-v_i^\text{ref}(r))-q_i\phi(r)+e_0\phi_i^\text{corr}(r)\right)\right)} \\
    & = l_i(r)/l_i^\text{ref}\; n_i^\text{ref} \exp{\left(\beta(\mu_i-\mu_i^\text{ref})-\beta q_i\phi(r)-\beta v_i^\text{ref}(r)\right)}
\end{align}
that produces an additional pre-factor $l_i(r)/l_i^\text{ref}$ to the MF result.

A similar calculation for the solvent density yields
\begin{align}
    n_s(r) &= 
     n_s^\text{ref} \exp\Bigg(\beta(\mu_s-\mu_s^\text{ref}) -\beta v_s^\text{ext}(r)+ \nonumber\\
     &+\log\Big(\frac{\sinh(p\beta\abs{\nabla\phi(r)})}{p\beta\abs{\nabla\phi(r)}}\Big)+\beta e_0\phi_s^\text{corr}(r)\Bigg) \\
    &= l_s(r)/l_s^\text{ref}\; n_s^\text{ref} \exp\Bigg(\beta(\mu_s-\mu_s^\text{ref}) + \nonumber\\
     &+\log\Big(\frac{\sinh(p\beta\abs{\nabla\phi(r)})}{p\beta\abs{\nabla\phi(r)}}\Big)-\beta v_s^\text{ext}(r)\Bigg)
\end{align}

It is evident that the LDA correlation functional introduces an activity coefficient for particle species $j$,
\begin{align}\label{1LActivityCoefficient}
    \gamma_j = (l_j(r)/l_j^\text{ref})^{-1} \ , 
\end{align}
which means the correlation potential can be rewritten as
\begin{align}\label{CorrelationPotential}
    \phi_j^\text{corr}(r) =  - (\beta e_0)^{-1} \log \left(\gamma_j\right). 
\end{align}
In bulk electrolytes, the activity coefficient is typically smaller than one,\cite{Debye_1923} indicating that $\phi_j^\text{corr}>0$ and consequently the density (for the same $\mu_j$) is larger compared to the MF case where $\gamma_j = 1$ and $\phi_j^\text{corr}=0$.

In summary, the variational derivative of the 1L LDA correlation functional with respect to $\phi$ corresponds to a correlation charge density, Eq.~\eqref{CorrelationAsChargeDensity}, that can be integrated into a redefined 1L permittivity Eq.~\eqref{1LDielectricConstant}. Additionally, the variational derivative of the 1L LDA correlation functional with respect to $n_j$ represents a correlation potential, Eq.~\eqref{CorrelationPotential} that depends solely on the species' local activity coefficients, $\gamma_j$, quantifying deviations from MF behavior. 
 
The derivation of the solution functional, Eq.~\eqref{ResultVariationalFunctional}, and its corresponding variational derivatives, Eqs.~\eqref{LDAVariationalDerivative1} and \eqref{LDAVariationalDerivative2}, constitute the primary technical result of this article. We have accomplished a transition from the thermodynamic partition function, Eq.~\eqref{GrandCanonicalPartitionFunction}, that describes an electrolyte solution in an external potential to a variational grand potential functional for particle densities and electrostatic potential, Eq.~\eqref{ResultVariationalFunctional}. The functional, when minimized, yields the thermodynamic equilibrium distributions of electrolyte densities and electric potential. We have derived a correlation functional, Eq.~\eqref{CorrelationFunctional}, that integrates coulombic correlation effects within the 1L and LDA. Via its functional derivatives, the 1L-LDA functional influences the Euler-Lagrange equations for the electric potential and electrolyte densities by introducing an additional correlation charge density and correlation potentials, respectively. 

\subsection{Applicability of the 1L-LDA functional}\label{SubSec:Applicability}
Derivation of the 1L-LDA functional involves two approximations. The key approximation in this article is the one-loop expansion, which allows accounting for coulombic correlation effects beyond the mean-field level. For an inhomogeneous system, i.e., in the presence of an external potential, the highest local ion concentration (density) determines the applicability limit of the 1L approximation. To assess the applicability of the 1L approximation, we evaluate the 1L approximation for a bulk electrolyte system, for which the LDA is exact. To derive an analytical expression for the upper bound of the density, the solvent is treated implicitly as a dielectric background $\epsilon$. By re-scaling in Eq.~\eqref{action}
\begin{align}
    \psi\to\frac{1}{\beta z e_0}\tilde{\psi},\quad
    r\to\lambda_D\tilde{r},
\end{align}
the action in Eq.~\eqref{GrandCanonicalPartitionFunction} becomes dimensionless (and independent of physical constants) and proportional to
\begin{align}\label{LoopParameter}
    \upsilon^{-1} = \frac{\lambda_D}{\lambda_B},
\end{align}
where $\lambda_B = (ze_0)^2/(4\pi\epsilon k_B T)$ is the Bjerrum length. This means that the value of the partition function Eq.~\eqref{GrandCanonicalPartitionFunction} is governed only by $\upsilon$. As $\upsilon$ becomes small, i.e., for low ion concentrations and large dielectric permittivity, the pre-factor of $S$ becomes large and the 1L approximation becomes more accurate.\cite{zinn-justinPhaseTransitionsRenormalisation2007} This defines our expansion parameter introduced in Eq.~\eqref{ExpansionAroundMeanField}.

In case of a charged metal--electrolyte interface, the highest ion concentration is achieved closest to the metal surface. In the classical Gouy-Chapman result\cite{schmicklerInterfacialElectrochemistry2010}, the sum of cation and anion densities at the metal surface as a function of the surface charge density, $\sigma_M$, is given by
\begin{align}\label{sumofDensity}
    n_c+n_a\approx\frac{\beta\sigma_{M}^{2}}{2\epsilon} \ .
\end{align}
Substituting Eq.~\eqref{sumofDensity} into the Debye length (Eq.~\eqref{DebyeLength}),  gives
\begin{align}\label{EffectiveDebyeLength}
    \lambda_D=\sqrt{2}\frac{\epsilon}{\beta ze_0 \sigma_M}.
\end{align}
Inserting Eq.~\eqref{EffectiveDebyeLength} into Eq.~\eqref{LoopParameter} yields the expansion parameter as a function of $\sigma_M$,
\begin{align}
     \upsilon = \frac{1}{\sqrt{2}}\frac{\beta^{2}(ze_{0})^{3}}{4\pi\epsilon^{2}}\sigma_{M}.
\end{align}
For given $\sigma_M$, we can thus estimate the applicability of the 1L expansion. Identifying the value of $\upsilon$ at which the 1L approximation becomes quantitatively inaccurate requires numerical methods, such as Monte Carlo simulations of the exact partition function.\cite{netzPoissonBoltzmannFluctuationEffects2000} For instance, Netz et al.\cite{netzPoissonBoltzmannFluctuationEffects2000} argued that the predictions of the 1L expansion become unphysical when $\upsilon>12$, although the 1L expansion can be quantitatively incorrect even for smaller values. For a 1:1 electrolyte, we set the upper bound conservatively to $\upsilon=1$, which gives an upper limit $\abs{\sigma_M}\approx4\,\mathrm{\mu Ccm^{-2}}$ or an ionic strength of $I_c=200\,\text{mM}$. We thus consider the 1L-LDA functional to be applicable around the potential of zero charge (pzc) as long as the electrode surface charge is smaller than $4\,\mathrm{\mu Ccm^{-2}}$.

\subsection{Coupling the electrolyte functional to other variational functional models}\label{SubSec:Embedding}
For the description of charged metal--electrolyte interfaces, the electrolyte model must be combined with a model for the metal electrode and surface charge. This is often achieved by heuristically adding up the different free energy contributions in a combined free energy functional.\cite{huangDensityPotentialFunctionalTheory2023} In such hybrid models, however, individual contributions from electrode and electrolyte subsystems and their interaction are not derived together from first principles. It is therefore not clear \textit{a priori} how the variational principle for the individual subsystems translates to the level of the combined free energy functional. The variational functional, as expressed in Eq. \eqref{ResultVariationalFunctional}, was derived under the presence of an external potential $v^\text{ext}_j$ for the electrolyte species. It will be shown in this section that the external potential naturally emerges from the coupling between different subsystems in a combined variational functional.

To get to this point, we define a combined grand potential functional as
\begin{align}\label{GrandPotentialFunctional}
    \Omega_\text{tot} = \mathcal{F}_\text{sol} + \mathcal{F}_\text{ext} - \sum_{j=a,c,s}\int_r \mu_j n_j(r),
\end{align}
which, compared to Eq.~\eqref{GrandPotentialSolutionFunctional}, has no external potentials but an additional free energy contribution, $\mathcal{F}_\text{ext}$. In classical density functional theory, external potentials and free energy functionals are used, for example, to model the influence of external charge distributions on the underlying statistical system. In this work, the external potential is used to model an explicitly considered additional subsystem, e.g. the metal electrode and its interaction with the electrolyte.

If we compare the variational derivatives of Eqs.~\eqref{GrandPotentialSolutionFunctional} and \eqref{GrandPotentialFunctional}, evaluated at the equilibrium density $n_j^{0}$ corresponding to $\mathcal{F}_\text{sol}$ for $v_j^\text{ext}=0$, we find
\begin{align}
    \frac{\delta \Omega_\text{tot}}{\delta n_j(r)} [n_j^0(r)] &= \frac{\delta \mathcal{F}_\text{ext}}{\delta n_j(r)} \label{Eq1}\\
    \frac{\delta \Omega_\text{sol}}{\delta n_j(r)} [n_j^0(r)] &= v_j^\text{ext}(r).\label{Eq2}
\end{align}
By comparing, Eqs.~\eqref{Eq1} and \eqref{Eq2} we arrive at the conclusion that any external functional acts upon the underlying statistical system in the same way as an external potential. 

The external functional represents the metal electrode and its interaction with the electrolyte, 
\begin{align}
    \mathcal{F_\text{ext}}=\Omega_\text{el}[n_e,\mu_e]+\mathcal{F}_\text{el-sol}[n_e,n_j],
\end{align}
where $\Omega_\text{el}$ is the electrode grand-potential functional that depends on the electron density, $n_e$, and the electron chemical potential, $\mu_e$, while being independent of the electrolyte densities. In the following we choose a simple MF coupling
\begin{align}\label{CouplingToExternalCharge}
    \mathcal{F}_\text{el-sol} = \int_{r,r'}\frac{\rho_\text{sol}(r)\rho_\text{el}(r')}{4\pi\epsilon_0\abs{r-r'}},
\end{align}
between the electrode charge $\rho_\text{el}(r)=(\rho_\text{core}(r)-e_0n_e(r))$, where $\rho_\text{core}$ represents the atomic core charge density of the metal, and the solution charge density
\begin{align}\label{SolutionChargeDensity}
    \rho_\text{sol}(r) = \sum_{i=a/c}q_in_i(r)+\nabla \epsilon(r)\cdot\nabla\phi(r),   
\end{align}
which consists of the charge due to ions and the bound charge due to the spatially dependent permittivity. It should be emphasized that this interaction functional between electrons and electrolyte corresponds to a MF coupling, i.e., it does not account for correlation effects between metal and electrolyte.\cite{binningerFirstprinciplesTheoryElectrochemical2023}

According to Eq.~\eqref{Eq1}, the external potential, which acts upon ions, is thus given by
\begin{align}
    v^{\text{ext}}_i (r) = q_i \int_{r'} \frac{\rho_\text{el}(r')}{4\pi\epsilon_0\abs{r-r'}} = q_i \phi_\text{el}.
\end{align}
Consequently, the external potential acting on species $i$ is the coulombic potential $\phi_\text{el}$ caused by $\rho_\text{el}$. As shown in Sec.~\ref{SubSec:FunctionalDerivative}, the Euler-Lagrange equation for $n_i$ in the presence of such an external potential reads
\begin{align}\label{DensityWithPhiExt}
    n_i(r) =  n_i^\text{ref} \exp{\left(\beta \left((\mu_i(r)-\mu_i^\text{ref })-q_i\phi_\text{tot} (r)+e_0\phi_i^\text{corr}(r)\right)\right)} \ ,
\end{align}
where the \emph{total} electrostatic potential combines the contributions from the electrode and electrolyte,
\begin{align}
    \phi_\text{tot} (r) = \phi(r)+\phi_\text{el}(r)\ .
\end{align}
By construction, $\phi_\text{tot}$ satisfies
\begin{align}
    \nabla(\epsilon_0\nabla\phi_\text{tot} (r)) = \nabla(\epsilon_0\nabla\phi(r))+\nabla(\epsilon_0\nabla\phi_\text{el}).
\end{align}
The second term on the r.h.s,
\begin{align}
    \nabla(\epsilon_0\nabla\phi_\text{el}(r)) &= \nabla\left(\epsilon_0\nabla\left(\int_{r'} \frac{\rho_\text{el}(r')}{4\pi\epsilon_0\abs{r-r'}}\right)\right) \nonumber \\
    &=\rho_\text{el}(r),
\end{align}
alongside the Poisson equation for $\phi$, as shown in Eq.~\eqref{VariationalEqPhi2}, we find the Poisson equation for $\phi_\text{tot}$,
\begin{align}\label{NewPoissonEquation1}
     -\nabla(\epsilon_0\nabla\phi_\text{tot}(r)) = \rho_\text{el}+ \rho_\text{sol}.
\end{align}
Hence, as expected, the Poisson equation for $\phi_\text{tot}$ includes the total charge density of the electrode--electrolyte system. We note that, vice versa, the electrolyte functional can be interpreted as an external potential for the metal-electronic subsystem leading to the same Poisson equation for $\phi_\text{tot}$. Accordingly, the electrostatic interactions within and across the entire system can be described by one common electrostatic potential $\phi_\text{tot}$, which, for the sake of simplicity, will be denoted as $\phi$ in the following. 

\section{Interface Model}\label{Sec: Application}
The second key goal of the article is to employ the developed functional formalism for studying correlation effects in the electrical double layer layer (EDL) and more specifically the influence of coulombic correlations on the interfacial capacitance. The electrode subsystem describes the metal and the electron density through an orbital-free DFT (OFDFT) model with a jellium background for the atomic charges. The electrolyte system describes the \textit{diffuse} part of the EDL and is modeled via the solution free energy functional (Eq. ~\eqref{ResultVariationalFunctional}) and an additional heuristic lattice gas model that describes volume exclusion.\cite{borukhovStericEffectsElectrolytes1997} The third subsystem is situated between the metal and the solution and represents a dense ionic layer. This contact interface layer is stabilized by an additional repulsive interaction that defines the distance of closest approach for the electrolyte species to the metal surface. The width and dielectric permittivity of the interface layer are adjustable such that the contact layer capacitance assumes a value such that the value of the total capacitance (that consists of the contact layer and diffuse layer capacitance) is in agreement with the capacitance value observed in the experimental data at the potential of zero charge (pzc). Keeping the contact layer capacitance constant, one can study the influence of the correlation functional on the properties of the diffuse part of the EDL. In the past, the combination of an OFDFT for the electrode together with a mean-field cDFT model for the electrolyte has been termed density-potential functional theory.\cite{huangGrandCanonicalModelElectrochemical2021} In this paper, instead of the MF functional, the 1L functional derived in Sec.~\ref{Sec: Theory} is used to model the electrolyte solution.

\subsection{Combined free energy 
functional for a metal-electrolyte system}\label{ACombinedFunctional}
\begin{figure}
	\centering
	\includegraphics[width=0.95\columnwidth]{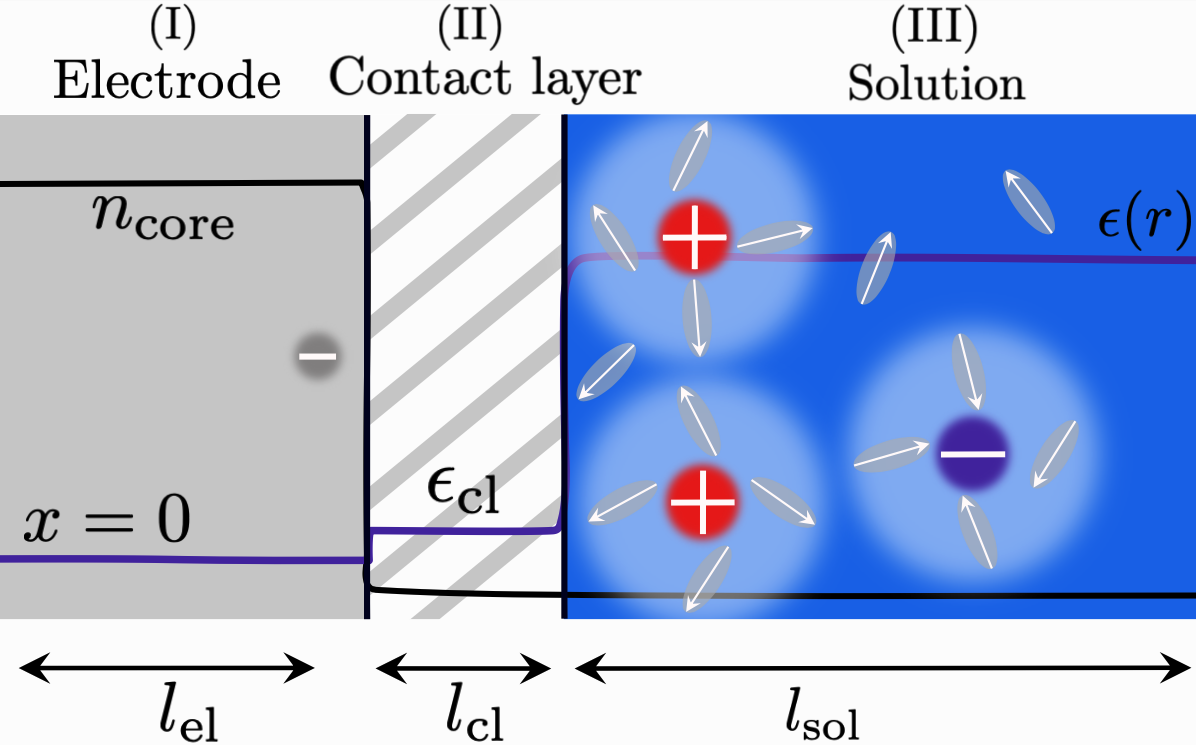}
	\caption{Schematic of the model region, consisting of three layers. The metal slab (layer I) of width $l_\text{el}$ is represented by a jellium model for the metal atomic cores with density $n_\text{core}$ (constant positive background) and electrons. It is in contact with an electrolyte layer (layer III) of width $l_\text{sol}$ consisting of hydrated cations (red), hydrated anions (blue), and solvent molecules (indicated by arrows). The latter contribute to the spatially dependent dielectric permittivity $\epsilon$ (schematically indicated as a purple curve). The electrode and the solution region are separated by a contact layer (layer II) of width $l_\text{cl}$ and dielectric permittivity $\epsilon_\text{cl}$. }
	\label{fig:conceptual}
\end{figure}

We here couple the OFDFT electrode model, represented by a free energy functional $\mathcal{F}_\text{el}$, to the electrolyte functional $\mathcal{F}_\text{sol}$ of Eq.~\eqref{ResultVariationalFunctional} via the interaction functional $\mathcal{F}_{\text{el-sol}}$ of Eq.~\eqref{CouplingToExternalCharge}. The combined grand potential functional of the electrode--electrolyte system thus reads
\begin{align}\label{GrandPotential}
\Omega[\phi,n_j,n_e]=&\left(\mathcal{F}_\text{el}-\int_r n_{e}\mu_{e}\right)+\mathcal{F}_{\text{el-sol}}+\mathcal{F}_\text{rep}+ \nonumber \\
&\left(\mathcal{F}_\text{sol}+\mathcal{F}_\text{st} -\sum_{j=a/c/s}\int_r n_{j}\mu_{j}\right),
\end{align}
where the round brackets contain contributions of electrode and electrolyte sub-systems, respectively. In addition to the fundamentally derived functional $\mathcal{F}_{\text{sol}}$, one empirical functional contribution was added to the solution system: the steric functional, $\mathcal{F}_{\text{st}}$ that embodies volume exclusion due to finite ion sizes as described by a lattice gas model. As shown in Fig.~\ref{fig:conceptual}, in between the electrode and the electrolyte, an additional interface contact layer was added through the heuristic repulsive functional, $\mathcal{F}_\text{rep}$, that prevents the electrolyte species from entering the metal electrode region. These additional functionals are explained in more detail in the following section. As derived in the previous section, the equilibrium distributions of the total electric potential, electron density, and electrolyte densities are obtained as solutions to the variational Euler-Lagrange equations of the combined grand potential functional, Eq.~\eqref{GrandPotential},
\begin{align}\label{VariationalEquation}
\frac{\delta\Omega}{\delta\phi}=0, \quad
\frac{\delta\Omega}{\delta n_{e}}=0, \quad
\frac{\delta\Omega}{\delta n_{j}}=0,  \quad j=a,c,s \ .
\end{align}
In the following, the additional functional contributions appearing in Eq.~\eqref{GrandPotential} will be introduced and the detailed set of Euler-Lagrange equations will be stated given in Eqs.~\eqref{ElectronDGL}, \eqref{ModifiedPoissonBoltzmann},  \eqref{ParticleDensities} and \eqref{SolventDensity}. 

\paragraph{Electrode} For more details of the OFDTF functional, we refer the reader to a previous publication.\cite{huangDensityPotentialFunctionalTheory2023} The OFDFT electrode functional consists of four terms,
\begin{align}\label{QuantumFreeEnergy}
\mathcal{F}_\text{el}=\int_r T_\text{e} + U_\text{ex}+U_\text{c}+U_\text{coul} \ ,
\end{align}
where
$T_\text{e}$ is the kinetic energy from Thomas-Fermi theory\cite{thomasCalculationAtomicFields1927}
\begin{align}\label{ThomasFermiTerm}
T_\text{e}= e_{au} a_0^{-3} t_\text{TF} (1+\theta_Ts^2),
\end{align}
with the atomic energy denoted by $e_{au} = e^2/(4\pi \epsilon_0 a_0)$ and the Bohr radius by $a_0$. The volumetric kinetic energy, \(t_\text{TF}(n_e)\), and correction $s(n_e,\nabla n_e)$ are functions of the electron density and its gradient. In our previous works the electrode model has been discussed in detail.\cite{huangDensityPotentialFunctionalTheory2023} Hence, only the physical meaning of the terms is provided herewith. For further details on the explicit dependence on electron density, please refer to the Appendix of \onlinecite{huangDensityPotentialFunctionalTheory2023}.  Exchange ($U_\text{ex}$) as well as correlation ($U_\text{c}$) terms are taken from the Perdew-Burke-Ernzerhof (PBE) functional,\cite{perdewGeneralizedGradientApproximation1996,thomasCalculationAtomicFields1927}
\begin{align}
	U_\text{ex}=e_{au} a_0^{-3} u_\text{ex} (1+\theta_\text{ex}s^2), \label{ExchangeTerm}\\
	U_\text{c}=e_{au} a_0^{-3} (u_\text{c}+\theta_c n_e a_0^3 t^2).\label{Correlation term}
\end{align}
In the above expressions, \(u_\text{ex}(n_e)\) is the volumetric exchange energy, and \(u_\text{c}(n_e)\) is an interpolation of the volumetric correlation energy of a uniform electron gas. The \(s(n_e,\nabla n_e)\), and $t(n_e,\nabla n_e)$ serve as corrections to the exchange correlation energies. While the parameters \(\theta_\text{ex}\), and \(\theta_\text{c}\) have recommended values, the parameter \(\theta_\text{T}\) needs to be determined for a specific metal. In a previous publication, \(\theta_\text{T}\) was determined for an Ag(111) electrode.\cite{huangDensityPotentialFunctionalTheory2023} This same electrode is modeled in this article . Finally, the electrostatic self-energy of the electrode subsystem is 
\begin{align}
    U_\text{coul}= \frac{1}{2}\int_{r,r'}\frac{\rho_\text{el}(r)\rho_\text{el}(r')}{4\pi\epsilon_0\abs{r-r'}} \ .
\end{align}
The respective Euler-Lagrange equations for the metal-electronic subsystem are obtained from the functional derivative of $\Omega$ with respect to $n_e$, cf. Eq.~\eqref{VariationalEquation}, yielding
\begin{align}\label{ElectronDGL}
a_{0}^{2}\nabla\cdot\nabla n_{e}=\frac{20}{3}n_{e}\frac{\omega}{\theta_{T}\omega-\theta_{X}'}\left(\frac{(\mu_{e}(r)-e_{0}\phi)-\mu_{e}}{e_{au}}\right),
\end{align}
with
\begin{align}
\mu_{e}(r)=e_{au}\left(a_{0}^{-3}\frac{\partial t_{TF}}{\partial n_{e}}+a_{0}^{-3}\frac{\partial u_{X}^{0}}{\partial n_{e}}+a_{0}^{-3}\frac{\partial u_{C}^{0}}{\partial n_{e}}\right).
\end{align}

\paragraph{Contact layer} The contact layer (layer II in Fig.~\ref{fig:conceptual}) serves two purposes. Firstly, the repulsive potential prevents the electrolyte from entering the metal skeleton. Secondly, it modifies the value of the total capacitance, similar to the Helmholtz capacitance in the GCS model. The contact layer is adjusted to match the value of the experimental capacitance at the pzc tuning $l_\text{cl}$ and $\epsilon_\text{cl}$. Once parameterized, the values of the contact layer properties are kept constant throughout this paper. Thus, the influence of the contact layer on the behavior of the \textit{diffuse} part of the EDL is kept constant.  The repulsive functional has the form
\begin{align}\label{InteractionFreeEnergy}
\mathcal{F}_{\text{rep}}=\int_{r}\sum_{j=a/c/s}n_{j}W_{j}
\end{align}
with functional derivatives
\begin{align}\label{VariationalEq4}
    \frac{\delta \mathcal{F}_{\text{rep}}}{\delta n_j(r)} = W(r)
\end{align}
where $W$ is the repulsive potential acting on electrolyte species for which we use
\begin{align}\label{LennardJones}
W(r)=\omega\cdot\Theta(l_\text{cl}-d(r)).
\end{align}
with $\omega>0$. The function $d(r)$ is the distance from position $r$ to the metal surface. For a planar interface, as depicted in Fig.~\ref{fig:conceptual}, with \(x\) being the coordinate perpendicular to the surface, \(d(r) = x - l_\text{el}\). As a consequence of the repulsive potential, within the region $l_\text{el}<x<l_\text{el}+l_\text{cl}$ is solvent free, resulting in a reduction of the permittivity to $\epsilon_\text{cl}$. The contact layer values $l_\text{cl}$ and $\epsilon_\text{cl}$ define the contact layer capacitance that is used in Sec.~\ref{SubSec:Interface} to align the total capacitance to the experimental value at the pzc. We fix parameters of the contact layer in all calculations in order to highlight the effect of the correlation functional on the properties of the \textit{diffuse} part of the EDL.

\paragraph{Electrolyte} The solution functional $\mathcal{F}_\text{sol}$ given in Eq.~\eqref{ResultVariationalFunctional} describes the diffuse layer of the EDL as discussed in Sec.~\ref{SubSec:FunctionalDerivative}. The steric functional $\mathcal{F}_\text{st}$ in Eq.~\eqref{GrandPotential} is responsible for finite-size effects of electrolyte species, which prevent unphysically large ion densities at the charged electrode--electrolyte interface. Following the Bikerman approach,\cite{bikermanStructureCapacityElectrical1942} steric effects are accounted for using the excess free energy of an ideal lattice gas.\cite{bikermanStructureCapacityElectrical1942, borukhovStericEffectsElectrolytes1997} There we  employ separate lattices for ions and solvent molecules, respectively. The ion lattice captures the lattice saturation effect at high surface charge densities, which, together with dielectric saturation, is responsible for the typical double-peak structure of electrochemical capacitance curves.\cite{borukhovStericEffectsElectrolytes1997} The solvent lattice serves to maintain the solvent density of the exp. relevant value.\cite{abrashkinDipolarPoissonBoltzmannEquation2007} In our previously published letter, we found the two-lattice approach necessary to obtain the correct value of the dielectric permittivity close to the metal surface.\cite{bruchIncorporatingElectrolyteCorrelation2024} In Bikerman theory, the free energy entails the steric contributions to the ion and solvent chemical potentials, given by
\begin{align}
\frac{\delta \mathcal{F}_\text{st}}{\delta n_i}&=\beta^{-1} \log \bigg(\frac{1}{1-n_{a}/n_{\text{ion,max}}-\alpha n_c/n_{\text{ion,max}}}\bigg),\quad \text{with }i\in a,c, \label{ExcessFreeEnergy1}\\
\frac{\delta \mathcal{F}_\text{st}}{\delta n_s}&=\beta^{-1} \log \bigg(\frac{1}{1-n_s/n_{\text{sol,max}}}\bigg),\label{ExcessFreeEnergy2}
\end{align}
where $n_\text{ion,max}$ and $n_\text{sol,max}$ are the maximum densities of the ionic and solvent lattices, respectively. Here, ions of unequal size are described by a relative size factor $\alpha=(d_c/d_a)^3$ multiplying the cation term in Eq.~\eqref{ExcessFreeEnergy1}, where $d_c$ and $d_a$ are the sizes of cation and anion. The anion size is simply given by $d_a=(n_\text{ion,max})^{-3}$, while the cation size is given by $d_c=(\alpha/n_\text{ion,max})^{-3}$. Similarly, the solvent size is $d_s=(n_\text{sol,max})^{-3}$. To summarize, the steric model has one free parameter for each of the three electrolyte species: $d_a,d_c$ and $d_s$. Alternatively, the maximum densities $n_\text{ion,max}$ and $n_\text{sol,max}$ can be specified, along with the relative size factor $\alpha$. From now on we stick to the former. 

Having introduced all contributions to the overall grand potential functional and their corresponding functional derivatives, we state the complete Euler-Lagrange equations for the electrolyte densities, $n_j$, and the electrostatic potential, $\phi$. As explained earlier, the Euler-Lagrange equation for $\phi$ is the Poisson equation with the total charge density of the system,
\begin{align}\label{ModifiedPoissonBoltzmann}
-\nabla(\epsilon(r)\nabla\phi)=\sum_{i=a/c}q_{i}n_{i}(r)+e_{0}(n_{cc}(r)-n_{e}(r)),
\end{align}
where the permittivity along the metal-electrolyte interface is given by
\begin{align}\label{EffectivePermittivity}
	\epsilon(r) = \begin{dcases}
1, &  x<l_\text{el}\\
\epsilon_\text{cl}, & l_\text{el}<x<l_\text{el}+l_\text{cl}\\
\text{Eq.~\eqref{1LDielectricConstant}}, & x>l_\text{el}+l_\text{cl}.
\end{dcases}
\end{align}
Inserting the results for the functional derivatives with respect to $n_j$ in Eqs.~\eqref{VariationalEq1}, \eqref{VariationalEq2},  \eqref{VariationalEq3}, \eqref{LDAVariationalDerivative2}, \eqref{VariationalEq4},\eqref{ExcessFreeEnergy1} and \eqref{ExcessFreeEnergy2} into \eqref{VariationalEquation} leads to the following expressions for the chemical potentials,
\begin{align}
\beta\mu_{i}&=\beta \mu_i^\text{ref}+\log\left(\frac{n_{i}(r)/n_i^\text{ref}}{l_{i}(r)/l_i^\text{ref}(1-n_{a}/n_{\text{ion,max}}-\alpha n_c/n_{\text{ion,max}})}\right) \label{LocalChemicalPotentialIons}\nonumber\\
&+\beta q_{i}\phi(r)+\beta W_{i}(r) \quad \text{for $i\in a/c$}, \\
\beta\mu_{s}&=\beta \mu_s^\text{ref}+\log\left(\frac{n_{s}(r)/n_s^\text{ref}}{l_{s}(r)/l_s^\text{ref}(1-n_{s}(r)/n_\text{sol,max})}\right) \nonumber\\
&-\log\left(\frac{\sinh p\beta\abs{\nabla\phi(r)}}{p\beta\abs{\nabla\phi(r)}}\right)+\beta W_{s}(r). \label{LocalChemicalPotentialSolvent}
\end{align}
The chemical potentials appearing on the left hand side of these expressions are determined by the (external) reservoirs for the grand-canonical ensemble and fix the respective bulk concentrations of ions (and solvent molecules) in the bulk of the electrolyte. We here note that in electrochemistry, chemical potentials \emph{including} electrostatic terms are typically referred to as \emph{electrochemical potentials} and denoted by $\tilde{\mu}_j$. In the present work, however, such a distinction between chemical and electrochemical potentials is unnecessary, because the fundamental chemical potentials $\mu_j$ introduced in the grand canonical partition function of Eq.~\eqref{Defn:PartitionFunction} naturally include all electrostatic contributions, i.e., they are electrochemical potentials by construction.

Until now we have not specified the reference state in the chemical potentials, \textit{cf.} Eq.~\eqref{LocalChemicalPotentialIons} and \eqref{LocalChemicalPotentialSolvent}. For a charged wall in contact with an electrolyte solution, the bulk acts as a reservoir for the solution species. Choosing the bulk electrolyte as the reference state $\mu_j^\text{ref}=\mu_j^b$, demanding chemical potential equilibrium $\mu_i=\mu_i^b$, and resolving Eqs.~\eqref{LocalChemicalPotentialIons} and \eqref{LocalChemicalPotentialSolvent} for the respective densities, leads to Boltzmann-like relations for $n_j$ as a function of $\phi$,
\begin{align}
n_{i}(r)&=n^\text{b}_\text{ion}\frac{l_{i}(r)}{l_{i}^{b}}\frac{\Theta_{i}(r)}{\mathcal{D}_\text{ion}(r)},\quad i\in a,c \label{ParticleDensities}\\
n_s(r) &= n^\text{b}_\text{sol}\frac{l_{s}(r)}{l_{s}^{b}}\frac{\Theta_{s}(r)}{\mathcal{D}_\text{sol}(r)}.\label{SolventDensity}
\end{align}
Here, the functions
\begin{align}
\mathcal{D}_\text{ion}(r)&=\chi_{\text{ion},v}+\frac{l_{a}(r)}{l_{a}^{b}}\chi_{a}\Theta_{a}(r)+\frac{l_{c}(r)}{l_{c}^{b}}\chi_{c}\Theta_{c}(r), \text{and} \\
\mathcal{D}_\text{sol}(r)&=\chi_{\text{sol},v}+l_{s}(r)/l_{s}^{b}\chi_{s}\Theta_{s}(r), 
\end{align}
embody steric effects. The parameters $\chi_j=n_j^bd_j^3$ are the bulk volume fractions of particle type $j$ normalized to maximum densities $d_j^{-3}$. Similarly, $\chi_{\text{ion},v}=(1-\chi_a-\chi_c)$ and $\chi_{\text{sol},v}=(1-\chi_s)$ are the bulk volume fractions of empty sites. The corresponding Boltzmann factors are
\begin{align}\label{BoltzmannFactor}
\Theta_{i}(r)&=\exp\left(-\beta W_{i}(r)-\beta q_{i}\phi(r)\right) ,\\
\Theta_{s}(r)&=\exp\left(-\beta W_{s}(r)+\log\left(\frac{\sinh(p\beta\abs{\nabla\phi(r)})}{p\beta\abs{\nabla\phi(r)}}\right)\right).
\end{align}
As explained earlier, the ratios $l_j(r)/l_j^b$ encode the impact of coulombic correlations on the particle densities in the form of effective activity coefficients. 

Eqs. \eqref{ElectronDGL} and \eqref{ModifiedPoissonBoltzmann}, together with the modified Boltzmann relations of Eqs.~\eqref{ParticleDensities} and \eqref{SolventDensity}, describe the self-consistent coupling of a metal density functional with an electrolyte density functional, with correlation effects included at the 1L level.

\subsection{Model parameters and boundary conditions}\label{SubSec:Parameters}
\begin{table}
    \caption{Parameters used in the MF and the 1L model for the EDL.}
	\begin{ruledtabular}
		\begin{tabular}{lll}\label{tab:NumericalParameters}
			Symbol & Description & Value\\
			\hline
            \multicolumn{3}{l}{Solution} \\
            $l_\text{sol}$ & Width of the solution region & $15\;\text{nm}$ \\
            $n_\text{ion}^\text{b}$& Ionic bulk concentration & $100\;\mathrm{mM}$\\
            $n_{s}^\text{b}$ & Solvent bulk concentration & $55.6\;\mathrm{M}$\\
			$d_a$ & Anion size & $11.2\;\text{\AA}$\\
			$d_c$ & Cation size & $12.3\;\text{\AA}$\\	
			$d_s$ & Solvent size &$3.1\;\text{\AA}$\\
			$z$ & Ionic charge number & $1$\\
            \multicolumn{3}{l}{Solution -- MF model} \\
			$p^\text{MF}$& MF water dipole moment&$ 4.8\;\mathrm{D}$\\		
            \multicolumn{3}{l}{Solution -- 1L model} \\
            $p^\text{1L}$& 1L water dipole moment&$ 1.8\;\mathrm{D}$\\	
            $a_s$& Solvent cutoff &$ 2.5\;\text{\AA}$\\
            $a_i\;(100\,\text{mM})$& Ionic cutoff &$ 3.23\;\mathrm{nm}$\\	
            $a_i\;(40\,\text{mM})$& Ionic cutoff &$ 3.75\;\mathrm{nm}$\\	
            $a_i\;(20\,\text{mM})$& Ionic cutoff &$ 4.73\;\mathrm{nm}$\\	
            \multicolumn{3}{l}{Contact-Layer} \\
            $l_\text{cl}$ & Width of the contact layer & $3\,$\AA\\
			$\omega$ & Potential energy & $\infty$\\
            $\epsilon_\text{cl}$ & Permittivity of the contact layer& $6.13\epsilon_0$\\ 
			\multicolumn{3}{l}{Electrode} \\
            $l_\text{el}$ & Thickness of the metallic substrate & $1\;\text{nm}$ \\
            $n_\text{cc}^0$ & Electron bulk density & $2.75\;\text{\AA}^{-3}$ \\
            $\theta_X$ & PBE parameter & $0.1235$ \\
            $\theta_C$ & PBE parameter & $0.046$ \\
            $\theta_T$ & Kinetic parameter & $2.08$ \\
		\end{tabular}
	\end{ruledtabular}
\end{table}

To investigate the influence of electrolyte correlations described by the 1L-LDA functional on EDL charging characteristics, we specifically aim to model the interface between an Ag(111) metal electrode and an aqueous $\mathrm{KPF_6}$ electrolyte solution. This particular system is chosen due to the weak specific adsorption of $\mathrm{PF_6^{-}}$ anions on Ag(111), rendering the system close to an ideal polarizable interface.\cite{huangDensityPotentialFunctionalTheory2023} Furthermore, experimental capacitance data over a wide range of electrolyte concentrations are available to validate and compare with the model.\cite{valetteDoubleLayerSilver1989} Table \ref{tab:NumericalParameters} provides a list of all free model parameters and the values chosen for them.

For a planar metal--electrolyte interface with translational symmetry in $yz$ direction, Eqs. \eqref{ElectronDGL} and \eqref{ModifiedPoissonBoltzmann} simplify to two coupled one-dimensional ordinary differential equations of the $x$ coordinate perpendicular to the interface. Figure \ref{fig:conceptual} illustrates a schematic of the model region. The dimensions of the electrode slab ($l_\text{el}=1\;\text{nm}$) and the electrolyte solution domains ($l_\text{sol}=15\;\text{nm}$) are sufficiently large to allow all fields and densities to reach their bulk values. In the bulk of the electrode, the electron density and the electric potential are uniform, corresponding to the boundary conditions
\begin{align}
\frac{\mathrm{d}n_{e}(x=0)}{\mathrm{d}x}&=0,\\
\frac{\mathrm{d}\phi(x=0)}{\mathrm{d}x}&=0.
\end{align}
At the opposite electrolyte boundary of the system (at $x=l_\text{el}+l_\text{cl}+l_\text{sol}$) the electron density is zero and, by definition, the electric potential reaches its zero reference value,
\begin{align}
n_{e}(x=l_\text{el}+l_\text{cl}+l_\text{sol})&=0,\\ \phi(x=l_\text{el}+l_\text{cl}+l_\text{sol})&=0.
\label{eq_electrolyte_potential_reference}
\end{align}

Two flavors of the combined free energy functional are compared to evaluate the influence of the correlation functional $\mathcal{F}^\text{corr,$1$L-LDA}_\text{sol}$ on the simulated structure and charging characteristics of the metal--electrolyte interface. The MF model corresponds to Eq.~\eqref{ResultVariationalFunctional} \emph{excluding} the correlation functional  $\mathcal{F}^\text{corr,$1$L-LDA}_\text{sol}$ , while the 1L model corresponds to the complete free energy functional of Eq.~\eqref{ResultVariationalFunctional}. Parameters specific to each model are indicated by the superscripts ``MF'' or ``1L'', listed in the respective sections in Table \ref{tab:NumericalParameters}. The equations for the MF model can be directly obtained by setting $l_j=1$ in the complete model.

\paragraph{Electrode} The metal parameters of the OFDFT model are adopted from the MF study previously presented for the Ag(111)-$\mathrm{KPF_6}$ interface.\cite{huangDensityPotentialFunctionalTheory2023} A uniform background charge $\rho_\text{core}(r)=e_0 n_{cc}(r)$, with
\begin{align}
n_{cc}(r)=n_\text{cc}^{0}\Theta(l_\text{M}-x) \ ,
\end{align}
represents the positive charge from the metal atom cores, where $n_\text{cc}^0=2.75\;\text{\AA}^{-3}$ is the effective density of the cationic metal cores that equals the electron density of cubic closed-packed Ag to maintain electroneutrality within the metal bulk. The parameters for the PBE functional are summarized in Tab.~\ref{tab:NumericalParameters}. Here, $\theta_X$ and $\theta_C$ have the recommended values while the gradient coefficient $\theta_T$ was determined by the MF study of the Ag(111)-$\mathrm{KPF_6}$ interface.\cite{huangDensityPotentialFunctionalTheory2023} As explained above, the chemical potentials used in this work correspond to the full \emph{electrochemical} potentials. Given the zero (electric) potential reference defined in the bulk of the electrolyte, the electron chemical potential, $\mu_e$, directly corresponds to the electrode potential $E=\mu_e/e_0$, which controls the surface charge of the metal electrode. In the following, we express the electrode potential as $E\text{ vs. } E_\text{pzc}$, where $E_\text{pzc}$ represents the potential of zero charge (pzc) of the electrode--electrolyte interface, i.e., the potential where the the metal surface carries zero electronic excess charge and there is zero net ionic charge accumulated in the EDL. On the pzc scale, a positive electrode potential thus indicates a positively charged metal surface, and vice versa.

\begin{figure*}[t]
	\centering
	\includegraphics[width=2\columnwidth]{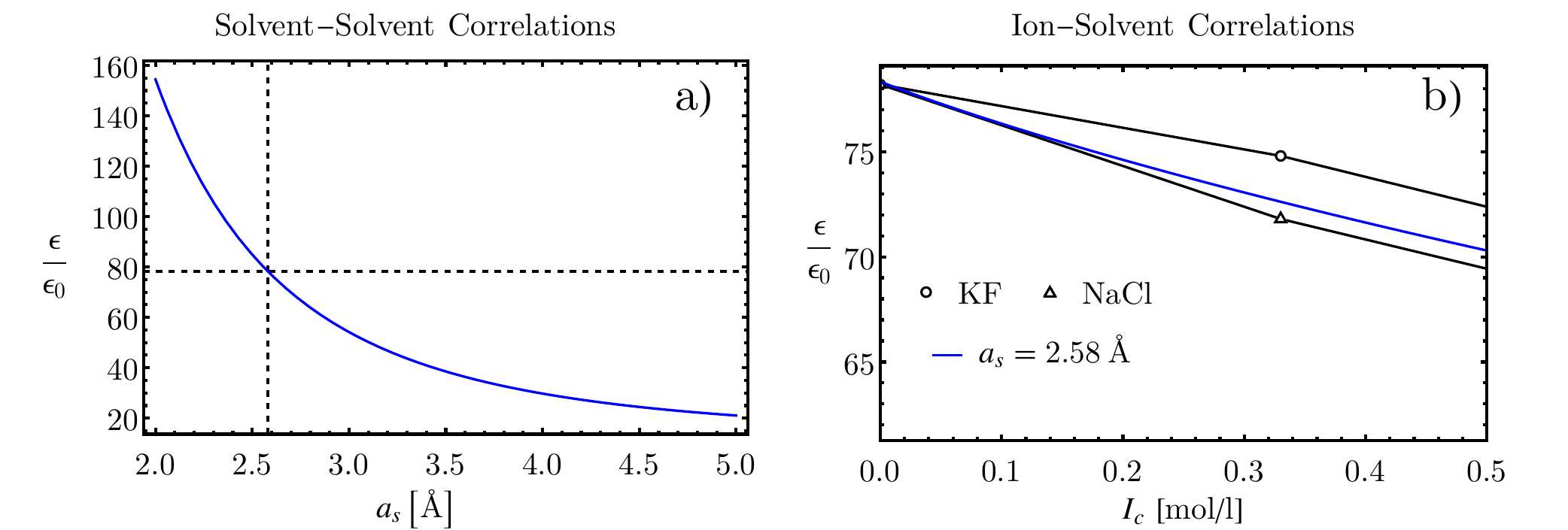}
	\caption{(a) The influence of solvent--solvent correlations on the dielectric permittivity of pure solvent (Eq.~\eqref{1LDielectricConstant}) is shown, as a function of $a_s$ for fixed $p^\text{1L}=p_\text{W}$. Two grid-lines indicate the experimental value $\epsilon_\text{W}$ on the y-axis and the corresponding value for $a_s$ on the x-axis. (b) Dielectric permittivity as a function of ionic strength $I_c$ representing the influence of ion--solvent correlations. Experimental values for NaCl (triangle) and KF (circles) for solutions up to $0.5\;\text{mol/l}$ concentration are shown\cite{haggisDielectricPropertiesWater1952} along with the model permittivity, Eq.~\eqref{1LDielectricConstant} for $a_s=2.58\,\text{\AA}$. }
 \label{fig:Calibration1}
\end{figure*}

\paragraph*{Electrolyte.} We next discuss the parametrization for the electrolyte subsystem. The sizes of the electrolyte species define the maximum density allowed by the lattice gas model. To model the incomprehensibility of water, we define the maximum density as the bulk water density $n_\text{sol,max}=55.6\;\mathrm{M}$,\cite{abrashkinDipolarPoissonBoltzmannEquation2007} which results in an effective size $d_s=3.1\;$\AA, which is slightly larger than the physical value $\approx2.7\,\text{\AA}$.\cite{svishchevStructureLiquidWater1993} As demonstrated by molecular dynamics (MD) simulations, the dimensions of the \(\text{PF}_6^-\) and \(\text{K}^+\) ions are $3-4\,\text{\AA}$ and $4-5\,\text{\AA}$, respectively.\cite{rajuAqueousSolutionBmim2009,rowleySolvationStructureNa2012} However, it is known that the sizes in the Bikerman model are effective fit parameters, which are often much larger than the physical values.\cite{bazantUnderstandingInducedchargeElectrokinetics2009} The results in Section \ref{SubSec:Interface} show that the model agrees well when anions have a radius of \(r_a=11.2\,\)\AA, while cations are slightly larger with a radius of \(r_c=12.3\,\text{\AA}\). Therefore, the trend that the cation is larger than the anion is consistent with MD simulations.\cite{rajuAqueousSolutionBmim2009,rowleySolvationStructureNa2012}

So far, the parameters have been equal in both the MF and 1L model. In the MF model, the permittivity Eq.~\eqref{MFDielectricConstant} depends only $p^\text{MF}$ and matches the experimental permittivity of bulk water $\epsilon^\text{W}=78.2\,\epsilon_0$ using $p^\text{MF}=4.8\,\mathrm{D}$, which is significantly larger than the actual water dipole moment of $p_\text{W}=1.8\;\text{D}$.\cite{abrashkinDipolarPoissonBoltzmannEquation2007,haggisDielectricPropertiesWater1952} In the 1L model, the permittivity Eq.~\eqref{1LDielectricConstant} is determined by two parameteres: $p^\text{1L}$ and the solvent cutoff $a_s$. This freedom allows to use the physical dipole moment of water $p^\text{1L}=p^\text{W}$ and determine $a_s=2.5\,\text{\AA}$ accordingly to match the experimental permittivity of bulk water. It was already highlighted in Eq.~\eqref{1LActivityCoefficient} that the parameter $a_i$ is related to the ionic activity coefficient. It was found that using $a_i=3.23\;\text{nm}$, Eq.~\eqref{1LActivityCoefficient} can qualitatively reproduce the experimental activity coefficient of KPF$_6$ as a function of ion concentration. The determination $a_s$ and $a_i$ though properties of a bulk electrolyte solution is a key result of this paper and is laid out in detail in Sec.~\ref{SubSec:Bulk}.

\paragraph{Contact layer} While the parameters for the electrode and electrolyte subsystem can be obtained independently of EDL properties, the contact layer region is added to repell the electrolyte from the interface and get agreement with experimental capacitance data at pzc. The repulsive potentials in Eq.~(\ref{LennardJones}) dictate the closest distance to the metal surface that the particles can reach. This distance is in principle arbitrary but is chosen to be $l_\text{cl}=3\;\text{\AA}$, which is range of typical simulation results for the metal-solution gap.\cite{sakongElectricDoubleLayer2018} As mentioned before, the contact layer permittivity $\epsilon_\text{cl}$ is a fit parameter for the EDL capacitance \textit{at pzc}. It was found that using $\epsilon_\text{cl}=6.13$, the capcacitance agrees with the experimental one at pzc. 

\section{Results}\label{Sec: Results}
The results section is divided into two parts. The first part explains the parametrization of the electrolyte functional of Eq.~\eqref{ResultVariationalFunctional} using experimental bulk electrolyte data for an aqueous KPF$_6$ solution, demonstrating that the correlation functional captures quantitative trends in the dielectric permittivity and ionic activity coefficient as functions of ionic concentration. The second part applies the fully parameterized EDL model of Eq.~\eqref{GrandPotential}, showing that including the correlation functional achieves quantitative agreement with the shape of the capacitance curve around the potential of zero charge (pzc). 

\subsection{Bulk electrolyte solutions}\label{SubSec:Bulk}
\paragraph*{Dielectric permittivity of pure solvent (water).}
In the MF model, the dielectric permittivity of a bulk solution ($\phi=0$), \textit{cf.} Eq.~\eqref{MFDielectricConstant}, can only be modified by tuning the MF dipole moment $p^\text{MF}$ assuming that bulk water concentration ($n_s^\text{b}=55.6\,\text{M}$) is fixed. The MF dielectric permittivity, $\epsilon^\text{MF}$, equals water permittivity, $\epsilon_\text{W}=78.2$,\cite{hastedDielectricPropertiesAqueous1948} only with a dipole moment $p^\text{MF}=4.8\;\text{D}$ that is significantly larger than the physical dipole moment of water $p_\text{W}=1.8\;\text{D}$. This discrepancy is a common limitation of the MF point-dipole model.\cite{abrashkinDipolarPoissonBoltzmannEquation2007} 

In the 1L model, solvent--solvent correlations contribute positively to the dielectric permittivity (Eq.~\ref{1LDielectricConstant}) of pure water. Therefore, in the 1L model, in addition to the dipole moment ($p^\text{1L}$), the solvent cutoff ($a_s$) can be tuned so that $\epsilon^\text{1L}=\epsilon_\text{W}$. Using the \textit{physical} dipole moment of water ($p^\text{1L}=p_\text{W}=1.8\;\text{D}$), the value of $\epsilon^\text{1L}$ is shown in Fig.~\ref{fig:Calibration1}a as a function of the solvent cutoff $a_s$. Solvent--solvent correlations are weakened, hence the permittivity is reduced by increasing the solvent cutoff. The required solvent cutoff to match the dielectric permittivity of pure water is about the size of a water molecule $a_s=2.58\,\text{\AA}$, which has been discussed in previous $1\mathrm{L}$ studies.\cite{levyDielectricConstantIonic2012} This shows that solvent--solvent correlation effects contribute significantly to the permittivity of pure water.

\paragraph*{Dielectric decrement as a function of ion concentration.} The MF dielectric permittivity $\epsilon^\text{MF}$ (Eq.~\eqref{MFDielectricConstant}) exhibits no dependency on the ionic concentration, contrary to experimental findings suggesting that the permittivity reduces with increasing salt concentration called dielectric decrement.\cite{levyDielectricConstantIonic2012} Two phenomena contribute to the dielectric decrement, first, the decrease of effective water concentration due to steric exclusion of water molecules by dissolved ions. Second, the loss of orientational degrees of freedom of water molecules in the solvation shell around dissolved ions. In the ion concentration range considered here ($<0.5\,\mathrm{mol/l}$), the former effect is negligible and the dielectric decrement thus mostly due to solvation.  The 1L model dielectric permittivity, $\epsilon^\text{1L}$, depends on the salt concentration via the Debye length $\lambda_D$ through the correlation parameter $l_s$ in Eq. \eqref{1LDielectricConstant}. Figure \ref{fig:Calibration1}b depicts $\epsilon^\text{1L}$ from Eq.~\eqref{1LDielectricConstant} as a function of ionic strength, $I_c=(n_\text{a}+n_\text{c})/2$, in comparison to experimental data of KF and NaCl. Since the solvent cutoff $a_s$ is already determined by the pure water permittivity, there is no free parameter left and the curve of $\epsilon^\text{1L}$ vs. $I_c$ is thus fixed. The respective plot in Fig.~\ref{fig:Calibration1}b shows a dielectric decrement with a slope that quantitatively well agrees to the experimental data, demonstrating that the 1L model is able to correctly capture the process of ion solvation in water in the form of a dipolar polarization shell around ionic point charges. 

The slope of the experimental dielectric decrement is salt ion species specific. In Section.~\ref{SubSec:Interface} an Ag(111)-KPF$_6$ interface is simulated for which the dielectric decrement of KPF$_6$ is of interest. Since experimental values for the dielectric decrement of KPF$_6$ are not available, the dielectric decrement of KPF$_6$ must be estimated. The dielectric decrement of a general salt can be decomposed into two distinct contributions, each due to one of the salt ion species.\cite{nakayamaNonlinearDielectricDecrement2023} The dielectric decrement due to K$^+$ is known,\cite{nakayamaNonlinearDielectricDecrement2023} in addition, larger anions have a stronger dielectric decrement, indicating that the total dielectric decrement of KPF$_6$ should be larger than KF but smaller than NaCl. This means the 1L model shown in Fig.~\ref{fig:Calibration1}b is in good quantitative agreement with the dielectric decrement estimated for KPF$_6$.

\begin{figure}
	\centering
	\includegraphics[width=1\columnwidth]{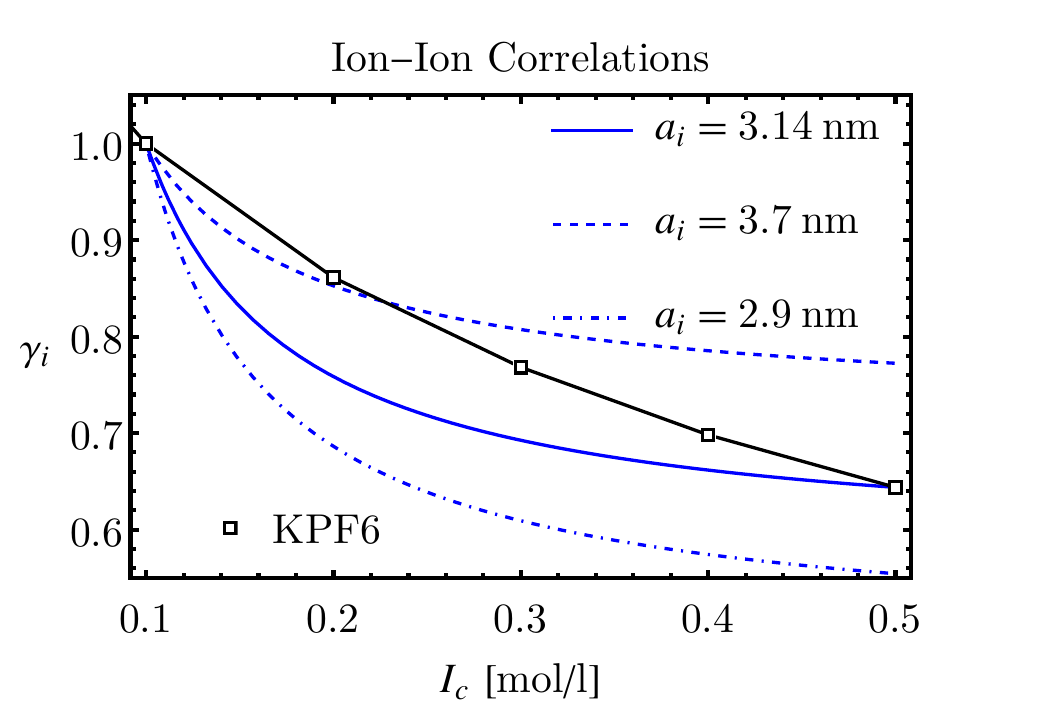}
	\caption{The activity coefficient, $\gamma_i$ (Eq.~\eqref{DHActivityCoefficient}), as a function of ionic strength $I_c$ for three different values for $a_i$. The experimental values of the mean activity coefficient of KPF$_6$(squares) are normalized to the respective value at  $I_c=100\;\text{mM}$.}
 \label{fig:Calibration2}
\end{figure}

\paragraph*{Ionic activity coefficient.}\label{SubSubSec:DeterminationLambdaI}

The ionic activity coefficient is a measure of the chemical-potential change arising from ion–ion correlations.\cite{bockrisModernElectrochemistryIonics1998} In the MF model, ion--ion correlations are absent and the the activity coefficient for all particle species is one. The activity coefficient in the 1L model is directly related to the inverse of the scaling factors $l_i$ in Eq.~\eqref{1LActivityCoefficient} with respect to a certain reference state. For the EDL model, it was highlighted in Sec.~\ref{ACombinedFunctional}, that it is convenient to choose the bulk reservoir of an electrolyte solution as the reference state, $\mu_i^\text{ref}=\mu_i^b$. W.r.t. the bulk reference, the activity coefficient, \textit{cf.} Eq.~\eqref{1LActivityCoefficient}, is given by
\begin{align}\label{DHActivityCoefficient}
    \gamma_i= (l_i/l^b_i)^{-1}.
\end{align}

\begin{figure*}
	\centering
	\includegraphics[width=2\columnwidth]{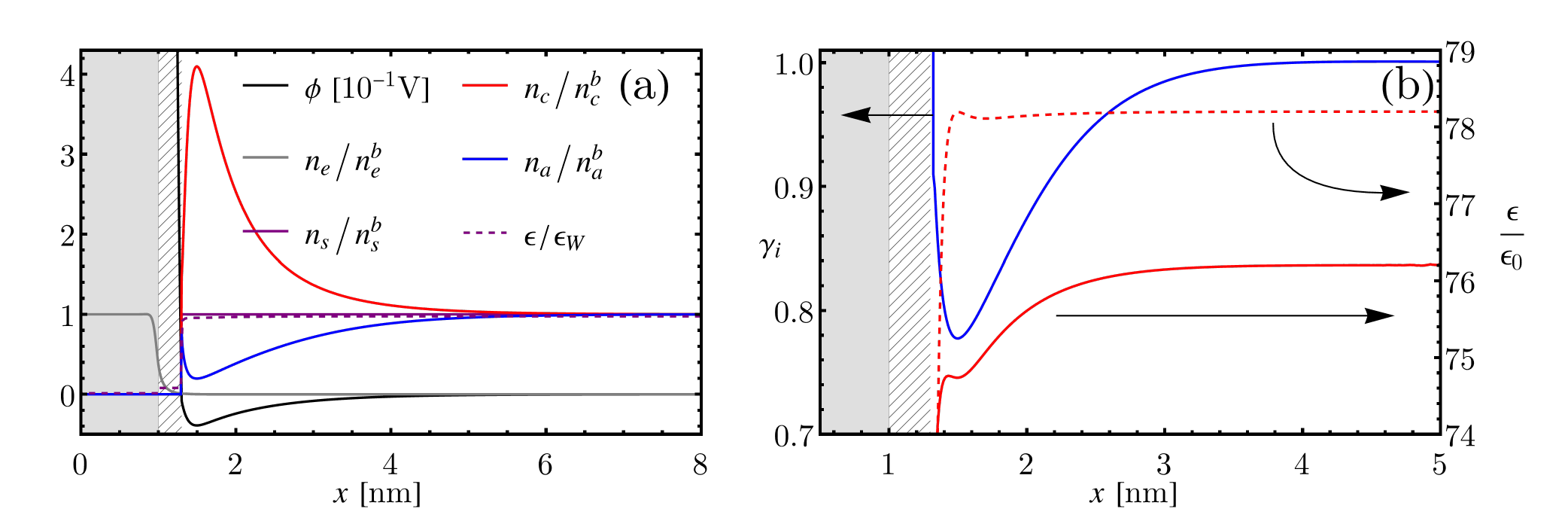}
	\caption{(a) Density and potential distributions at the electrode--electrolyte interface obtained from the 1L model with parameters according to Tab.~\ref{tab:NumericalParameters} at $E=-0.1\,\text{V}\,\text{vs}\,E_\text{pzc}$ and a $100\,$mM bulk electrolyte concentration. The gray area denotes the electrode and the hatched area denotes the contact layer region. (b) The local activity coefficient (blue) and the dielectric permittivity according to the MF (red dashed) and the 1L model (red solid) for the same parameters as in (a).}
 \label{fig:1LSimulations}
\end{figure*}

The activity coefficient as a function of ionic strength is shown in Fig. ~\ref{fig:Calibration2} for three different values of the parameter $a_i$, using a reference state of $100\,$mM ionic strength. The model curves are compared with experimental data of the mean activity coefficient ($\gamma_\pm$) of an aqueous KPF$_6$ solution. For a $1:1$ electrolyte solution, the mean activity coefficient $\gamma_\pm$ is equal to the individual ionic activity coefficient $\gamma_\pm=(\gamma_a\gamma_c)^{1/2}=\gamma_i$.\cite{bockrisModernElectrochemistryIonics1998} In order to compare the 1L model with experiment, it is essential that all activity coefficients are expressed with respect to the same reference state. Experimental mean activity values in Fig.~\ref{fig:Calibration2} are normalized to the mean activity coefficient at  $100\,$mM ionic strength.\cite{chemistryQuantitiesUnitsSymbols1993}

The ionic cutoff, $a_i$, strongly affects the slope of the activity curve in Fig.~\ref{fig:Calibration2}. The optimal value of $a_i$ depends on the concentration range considered. In Sec.~\ref{SubSec:Interface}, the 1L-LDA functional is applied to the study of an electrode--electrolyte interface where the maximum interfacial ion concentration is about $500\,\text{mM}$ in the range of considered electrode potential. Accordingly, the value of $a_i=3.14\,\text{nm}$ is chosen to match the value of the 1L activity coefficient ($\gamma_i$) to the experimental one at an ionic strength of $I_c=500\,\text{mM}$. 

It is observed in Fig.~\ref{fig:Calibration2} that the activity coefficient of the 1L model declines too steeply at low concentration and too weakly at higher concentration in comparison to experiment. This discrepancy is mainly due to the fact that the 1L equation for the activity coefficient (Eq.~\eqref{DHActivityCoefficient}) effectively contains the MF expression for the dielectric permittivity (Eq.~\eqref{MFDielectricConstant}). As discussed earlier, in the 1L model, the physical dipole moment of the water molecule is employed, which yields the correct dielectric permittivity at the 1L level, but results in a significantly lower permittivity when used in the MF expression. The result is an effective Debye length that is considerably smaller than the value obtained when the true bulk water dielectric constant used. Using a larger dipole moment, closer to the MF dipole moment $p^\text{1L}\approx p^\text{MF}$, results in a more accurate Debye length and thus a better agreement with the activity coefficient. However, this also causes the dielectric decrement to disappear. At the 1L stage of the model, it is thus not possible to improve the quantitative agreement for the activity coefficient without worsening the agreement for the dielectric decrement. Further development going beyond the 1L approximation, will be needed to resolve this ambiguity. 

It should be noted that the discrepancy between the effective Debye length and the physical one is confined to this particular aspect of the activity coefficient. The Poisson equation contains the 1L dielectric permittivity, \textit{cf.} Eq.~\eqref{PoissonEq1L}, and thus provides the correct Debye length. Therefore, quantities on a 1L level such as the interfacial capacitance, calculated in Sec.~\ref{SubSec:Interface}, are not influenced by the relatively small effective Debye length.

Depending on the bulk electrolyte concentration considered, the reference state for the activity coefficient changes, and the value of $a_i$ must be adjusted accordingly. Tab.~\ref{tab:NumericalParameters} displays values of $a_i$ for the bulk concentrations $40\;\text{mM}$ and $20\;\text{mM}$. For smaller bulk concentrations, the required value of $a_i$ increases. 

In summary, applied to the bulk electrolyte solution, the 1L-LDA functional correctly captures solvent--solvent correlations that make a significant contribution to the bulk water permittivity. The obtained dielectric decrement due to the solvation of ions (solvent--ion correlations) is also in good agreement with the experimental data. Ion--ion correlations are correctly found to reduce the ionic activity coefficient which qualitatively agrees with the experimental activity coefficient for KPF$_6$ solutions, although certain quantitative discrepancies exist.

\subsection{Metal--Solution interface}\label{SubSec:Interface}

The parametrization of the 1L-LDA functional was fixed as described in Sec.~\ref{SubSec:Bulk}. In this section, we investigate the role of \textit{bulk} electrolyte correlation effects, captured by the parametrized 1L-LDA functional, on interfacial EDL properties. For the sake of clarity, when utilizing the 1L-LDA functional, fields and parameters are denoted by the superscript ``1L'', while for the MF model the superscript ``MF'' is used. The respective parameters are discussed in Sec.~\ref{SubSec:Parameters}.

\begin{figure*}
	\centering
	\includegraphics[width=2\columnwidth]{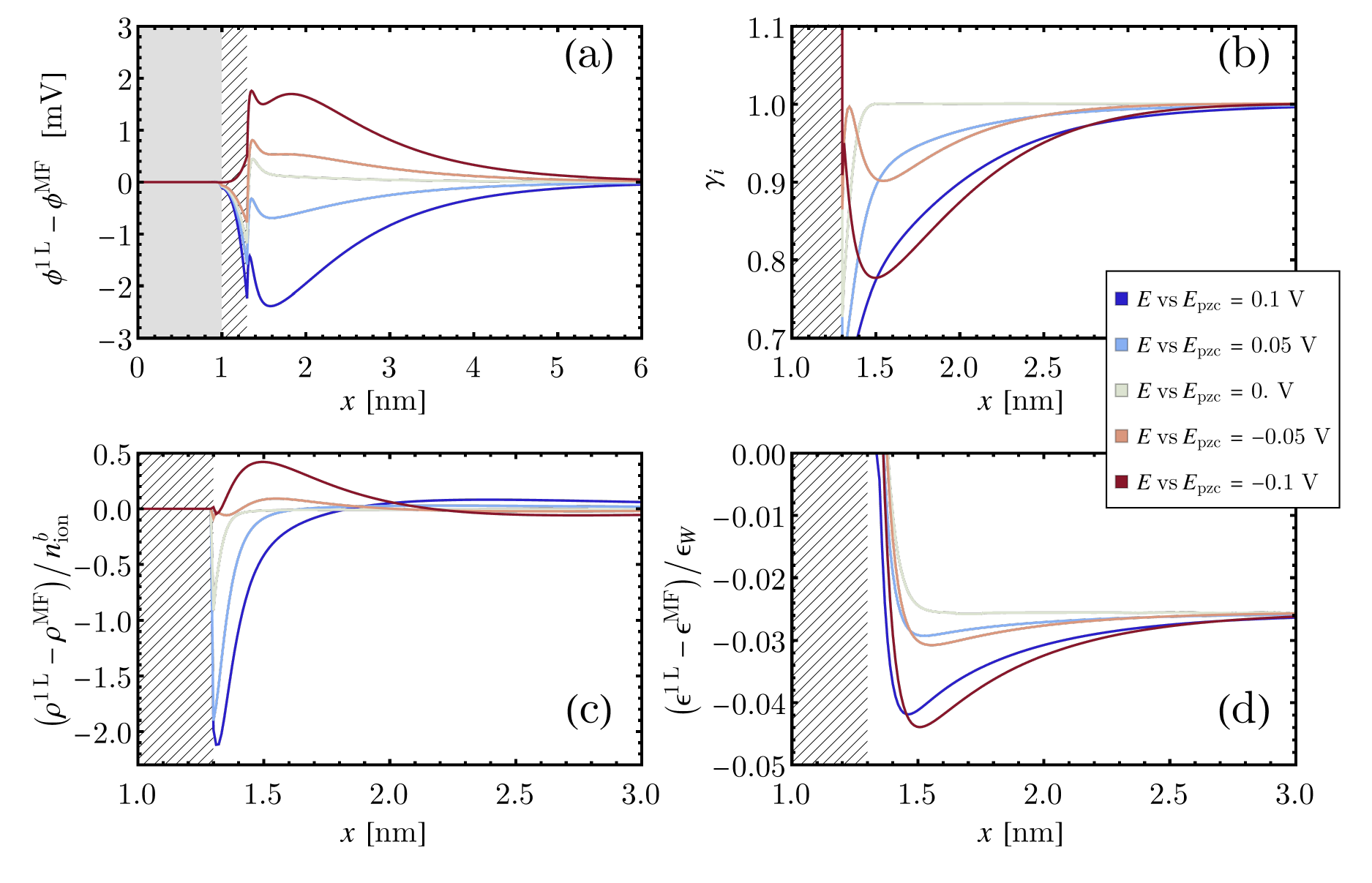}
	\caption{This figure shows from $E\,\text{vs}\,E_\text{pzc}=0.1\,\text{V}$ to $E\,\text{vs}\,E_\text{pzc}=-0.1\,\text{V}$ (a) the electric potential corrections in $\text{[mV]}$, together with (b) the activity coefficient for ions normalized to the bulk value and (c) the change in excess charge density of the electrolyte normalized to the bulk concentration. In (d), the change in dielectric permittivity normalized to bulk water permittivity is shown.}
	\label{fig:1LoopCorrectionsSummary}
\end{figure*}

\paragraph*{Dielectric permittivity and activity coefficient in the EDL.} By shifting to a positive or negative electrode potential with respect to the pzc, the electrode is charged with a positive or negative excess of electronic charge, respectively. In Fig.~\ref{fig:1LSimulations}a, the electrostatic potential $(\phi)$, the electrolyte densities ($n_a,n_c,n_s$), and the dielectric permittivity ($\epsilon$) are shown as functions of the coordinate $x$ perpendicular to the electrode surface for a negative electrode potential of $E=-0.1\,\text{V}\,\text{vs}\,E_\text{pzc}$. Due to the negative surface charge, the electrostatic potential in the diffuse layer is negative, which raises the cation density and reduces the anion density relative to the electrolyte bulk value, see Eq.~\eqref{ParticleDensities}. The dielectric permittivity closely follows the solvent density to which it is proportional (Eq.~\eqref{EffectivePermittivity}).

In Sec.~\ref{SubSec:Bulk} it was shown that electrolyte correlation effects become manifest in the dielectric permittivity and activity coefficient. The present formalism allows to plot the \emph{local} dielectric permittivity (Eq.~\eqref{EffectivePermittivity}) and \emph{local} activity coefficient (Eq.~\eqref{DHActivityCoefficient}) across the EDL as functions of  $x$, shown in Fig.~\ref{fig:1LSimulations}b. Away from the electrode, where the densities are uniform, the activity coefficient approaches $\gamma_i=1$, since the bulk electrolyte is defined as the reference state for the activity coefficient, see Eq.~\eqref{DHActivityCoefficient}. Closer to the interface, the local ionic strength increases, resulting in a \textit{lower} activity coefficient, \textit{cf.} also Fig.~\ref{fig:Calibration2}. An activity coefficient smaller than one means that the the ion density (Eq.~\eqref{ParticleDensities}) is multiplied by a scaling factor ($\gamma_i^{-1}=l_i/l_i^b$) which is larger than one. In other words, the ion density is \textit{enhanced} compared to the MF case, where the activity coefficient has a value of one. 

\begin{figure*}
	\centering
	\includegraphics[width=2\columnwidth]{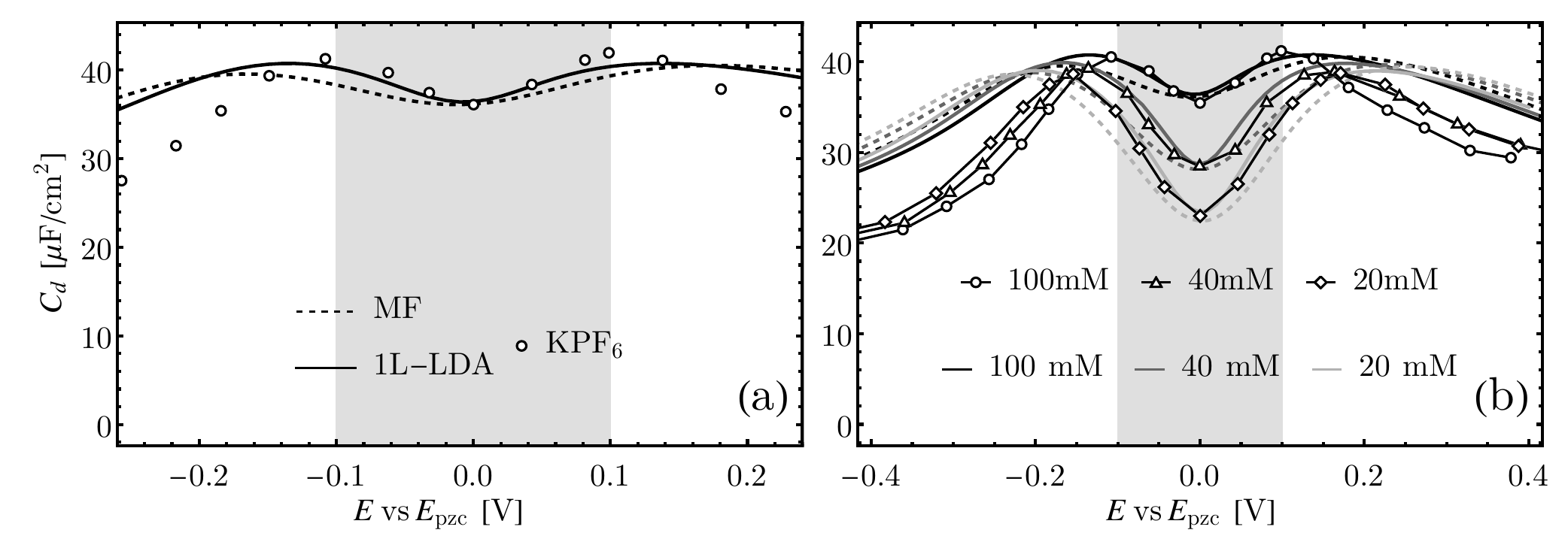}
	\caption{(a) The MF and the 1L model are compared with experimental capacitance data at 100mM bulk concentration. (b) The LDA simulations are compared with experimental values at three different bulk ion concentrations. The parameters used are shown in Tab. \ref{tab:NumericalParameters}. The estimated range of validity for the 1L expansion is highlighted in light-gray. All experimental curves are rescaled by the roughness factor (1.08) provided by Valette.\cite{valetteDoubleLayerSilver1989,daikhinDoublelayerCapacitanceRough1996}}
	\label{fig:ComparisonToExp}
\end{figure*}

In the MF model, the dielectric permittivity (red dashed curve in Fig.~\ref{fig:1LSimulations}b) assumes the value of pure water in the electrolyte bulk, as already discussed in Sec.~\ref{SubSec:Bulk}. The strong reduction in dielectric permittivity, close to the interface in Fig.~\ref{fig:1LSimulations}b, is determined by two competing effects. On the one hand, the dipolar solvent medium is attracted to the interface, due to the interfacial electric field, causing an increased solvent density and, hence, an increase in dielectric permittivity. On the other hand, the same interfacial electric field reduces the dielectric permittivity due to dielectric saturation, \textit{cf.} the Langevin function ($\mathcal{L}$) in Eq.~\eqref{MFDielectricConstant}.\cite{abrashkinDipolarPoissonBoltzmannEquation2007} In the 1L model, the value of the dielectric permittivity in the electrolyte bulk (red solid curve in Fig.~\ref{fig:1LSimulations}b) is lower than that of pure water due to the solvation effect discussed in Sec.~\ref{SubSec:Bulk}. Close to the electrode surface, the accumulation of charge leads to an increase in local ionic strength, from $I_c=100\,\text{mM}$ in the electrolyte bulk, to $I_c\approx150\,\text{mM}$, \textit{cf.} Fig.~\ref{fig:1LSimulations}a. The increase in ionic strength in the diffuse layer reduces the dielectric permittivity in the region $1\,\text{nm}$ away from the contact layer by about $2\%$ relative to the bulk value. In contrast, the variation of the activity coefficient from the electrolyte bulk is approx. $20\%$. It can be concluded that ion--ion correlations are considerably more important than ion--solvent correlations around the pzc. 

Subsequently, correlation effects are evaluated as a function of electrode potential. In Figure~\ref{fig:1LoopCorrectionsSummary}, 1L model predictions are  compared directly to MF model predictions of the electrostatic potential $\phi$, the activity coefficient $\gamma_i$, the excess charge $\rho$ and the dielectric permittivity $\epsilon$ using the parameters from Table~\ref{tab:NumericalParameters}. Two effects influence the counterion density in the EDL, \textit{cf.} Eq.~\eqref{ParticleDensities}, namely the electrostatic potential and the activity coefficient. Figure~\ref{fig:1LoopCorrectionsSummary}a shows that the electrostatic potential $\phi$ of the 1L model increases for negative $E$ relative to the pzc. For negative $E$, the cations are the counterions. Hence, if only the electrostatic potential would be affected, the cation density would be reduced compared to the MF model. However, Fig.~\ref{fig:1LoopCorrectionsSummary}b shows that the activity coefficient in the EDL is decreasing for larger $E$. This is entirely attributed to the fact that at higher $E$, the ionic strength in the EDL is elevated. The decrease in $\gamma_i$ alone would result in an increase in cation density, \textit{cf.} Eq.~\eqref{ParticleDensities}. The excess charge density, $\rho(r)$, of the electrolyte,
\begin{equation}
\rho(r)=n_c(r)-n_a(r),
\end{equation}
is shown in Fig.~\ref{fig:1LoopCorrectionsSummary}c. In total, the reduction in activity coefficient (Figure~\ref{fig:1LoopCorrectionsSummary}b) more than compensates the elevation of the electrostatic potential (Figure~\ref{fig:1LoopCorrectionsSummary}a), resulting in a net increase in counterion density. 

In general, it can be observed that the ion--ion correlation-induced reduction in local activity coefficient results in increased counterion density at the interface when compared to the MF model, for a given value of $E\,\text{vs.}\,E_\text{pzc}$. The physical reason for this behavior is that the MF model overestimates the Coulomb repulsion between like-charged counterions.\cite{netzPoissonBoltzmannFluctuationEffects2000} By accounting for screening of coulombic interactions, as discussed in Sec.~\ref{SubSec:Bulk}, the mutual repulsion of like-charged particles is reduced, which in turn increases the counterion density. 

The curves for positive and negative $E$ (relative to $E_\text{pzc}$) are slightly asymmetrical about the abscissa. The asymmetry is attributed to electronic and finite size effects that affect cation and anion species differently. Correlation effects, on the other hand, embodied in the scaling factors $l_i$, \textit{cf.} Eq.~\eqref{ParticleDensities}, are affecting both ion species on equal footing.

The difference in dielectric permittivity computed from the MF and 1L model, shown in Fig. 
\ref{fig:1LoopCorrectionsSummary}d, is slightly negative in the electrolyte bulk consistent with the difference shown in Fig.~\ref{fig:1LSimulations}b. For larger electrode potentials the ionic strength at the interface is higher. Due to solvation, in the 1L model, the dielectric permittivity at the interface is smaller for larger electrode potentials, compared to the MF model, which does not capture solvation effects. Overall, the value of the dielectric permittivity in the diffuse layer is reduced relative to the MF case for all electrode potentials considered in this study, and further decreases at the interface with larger $E$. However, the dielectric permittivity decrease is about $5\,\%$ of the bulk water value around the electrode potential window considered here (around the pzc).

\paragraph*{Interfacial capacitance.} The interfacial capacitance is an important footprint of electrified interfaces and experimentally accessible. The differential capacitance can be calculated from the surface charge density,
\begin{align}
    \sigma_M=e\int_0^{\infty} \left(n_{\text{cc}}(x)-n_e(x) \right)dx,
\end{align}
as the partial derivative with respect to $E$,
\begin{equation}
C_d=\frac{\partial \sigma_M}{\partial E}.
\end{equation}

In Fig.~\ref{fig:ComparisonToExp}a, the differential capacitance is shown for the parameters presented in Tab.~\ref{tab:NumericalParameters} along with experimental capacitance data of an Ag(111)-KPF$_6$ system (circles).\cite{valetteDoubleLayerSilver1989} In grey, the electrode potential range is highlighted where the 1L-LDA functional is assumed to be accurate, as the surface charge density does not exceed the upper bound estimated in Sec.~\ref{SubSec:Applicability}.  In the MF model (dashed), where the 1L-LDA functional is not used, the double-peak structure of the capacitance curve is reproduced, though there are quantitative discrepancies in peak height and peak-to-peak distance compared to experiment.  Using the 1L-LDA functional, the capacitance peaks are significantly more pronounced, and the peak-to-peak distance is smaller compared to the MF model. The more pronounced shape is due to the elevated excess charge at equal electrode potential, which was observed in Fig.~\ref{fig:1LoopCorrectionsSummary}c. The reduced activity coefficients at the interface (see Fig.~\ref{fig:1LSimulations}b), increase the counterion density and thus lead to an increase in surface charge density at the same electrode potential. The reduction in dielectric permittivity, \textit{cf.} Fig.~\ref{fig:1LoopCorrectionsSummary}d, partially counterbalances the increase of interfacial capacitance. It is noteworthy that using the 1L-LDA functional significantly improves the agreement with the experimental data around the pzc, which is a key finding of this article. At higher electrode potentials $( \vert E\,\text{vs.}\,E_\text{pzc}\vert > 0.1\,\text{V})$, differential capacitance curves computed from the MF and the 1L model  overestimate experimental values. In this potential range, steric effects dominate the EDL structure, which are accounted for in this work by the heuristic lattice gas model.\cite{bikermanStructureCapacityElectrical1942} However, the 1L model produces lower capacitance values compared to the MF model. The lower capacitance is due to lower dielectric permittivity in the 1L model, which results in a lower diffuse layer capacitance. Model results for concentrations of $100\;\text{mM}$, $40\;\text{mM}$, and $20\;\text{mM}$ (solid), along with experimental results (symbols), reported by Valette,\cite{valetteDoubleLayerSilver1989} are presented in Fig.~\ref{fig:ComparisonToExp}b. Additionally, the MF model results are presented in the same color but dashed. In the region around the pzc, where the 1L-LDA functional applies, which is the potential range $\vert E\;\text{vs.}\;E_\text{pzc}\vert<0.1\;\text{V}$, the model agrees well with experimental results across all electrolyte concentrations. This shows that electrolyte correlation effects are important for the characteristics of the differential capacitance, even at dilute solutions. 

\subsection{Conclusions}\label{Sec:Discussion}
This article has presented the derivation of a free energy functional for an electrolyte solution based on effective field theory. This approach allows including coulombic correlations with a 1L expansion, thus overcoming limitations of common MF approaches. The two essential steps involve a transformation to a variational functional for the electrostatic potential using an auxiliary charge and a transformation to a variational functional for the electrolyte densities using auxiliary potentials. The key theoretical contribution is the derived variational functional that separates into an ideal gas part and two excess parts. One excess part is the common MF functional and the other is a 1L-LDA correlation functional. The correlation functional captures coulombic correlation effects of an electrolyte solution. In this article, the correlation functional parameters are calibrated using experimental electrolyte solution data. 

The 1L-LDA correlation functional correctly reproduces experimental trends in dielectric permittivity and ionic activity coefficient, which are known to be affected by coulombic correlation effects.\cite{adarDielectricConstantIonic2018,Debye_1923} The 1L model contains solvent--solvent correlations that affect the dielectric permittivity of bulk water. Furthermore, the 1L model describes ion--solvent correlation effects and as such describes solvation of ions in water that reduce the dielectric permittivity as a function of ionic strength. Similarly, the 1L model includes ion--ion correlations, which reduce the activity coefficient upon increasing ionic strength. Embedded into an EDL functional, the reduction of the activity coefficient at the interface that is induced by ion--ion correlations increases the counterion density at a given electrode potential. The 1L-LDA functional achieves significantly improved agreement with experimental data, compared to the MF prediction around the pzc. This agreement exists over a wide concentration range. In the future, the goal is to move beyond the 1L approximation by systematically including higher-order terms, extending the applicability of the approach to higher concentrations.

\begin{acknowledgments}
	The work was carried out within the framework of the Helmholtz Program Materials and Technologies for the Energy Transition in the topic Chemical Energy Carriers.	
\end{acknowledgments}

\section*{Author Declarations}
\subsection*{Conflict of Interest}
The authors have no conflicts to disclose.

\subsection*{Author Contributions}
\textbf{Nils Bruch: }Conceptualization, Methodology, Formal analysis, Writing – original draft, Writing – review \& editing. \textbf{Tobias Binninger: }Conceptualization, Methodology, Formal analysis, Writing – original draft, Writing – review \& editing. \textbf{Jun Huang: }Conceptualization, Writing – review \& editing. \textbf{Michael Eikerling: }Conceptualization, Writing – review \& editing.

\section*{Data Availability Statement}
The data that support the findings of
this study are available from the
corresponding author upon reasonable
request.

\appendix
\section{Calculations for the LDA correlation functional}
\subsection{Calculation of the second variation in Eq.~\eqref{InverseGreen}}\label{Derivation of the second variation}
The first variational derivative of the action with respect to the field $\psi$ is
\begin{align}\label{FirstVariationOfAction}
\frac{\delta S[\psi]}{\delta\psi(r)}&=\sum_{i=a/c}iq_{i}\lambda_{i}\Lambda_{i}^{-3}e^{-iq_{i}\beta\psi(r)}\nonumber 
\\-\bigg(\epsilon_0+\lambda_{s}&\Lambda_{s}^{-3}p^{2}\beta\frac{\sinh(u)}{u}\left(\mathcal{L}^{2}(u)+\mathcal{L}'(u)\right)\bigg)\nabla^{2}\psi(r),
\end{align}
where, for convenience, we repeat omitting the argument, $\mathcal{L}\equiv\mathcal{L}(u)$ and $\mathcal{L}'\equiv\mathcal{L}'(u)$ are the Langevin function and its derivative, respectively, and $u=p\beta \vert \nabla \psi \vert$. The second variation can be obtained using the functional derivative on Eq.~(\ref{FirstVariationOfAction}). Some useful relations in these calculations are
\begin{align}
&\frac{d}{dx}\left(\frac{\sinh(u(x))}{u(x)}\right)=\frac{\sinh(u)}{u}\mathcal{L}(u)\frac{du}{dx}, \nonumber\\
&\frac{\partial}{\partial\nabla\psi}\abs{\nabla\psi}=\frac{\nabla\psi}{\abs{\nabla\psi}},\nonumber\\
&\nabla\abs{\nabla\psi}=\frac{\nabla\psi}{\abs{\nabla\psi}}\nabla^{2}\psi,\quad\text{and}\quad \nabla\frac{\nabla\psi}{\abs{\nabla\psi}}=0.
\end{align}
To obtain the second variational derivative, one can rewrite the first variation, Eq.~\eqref{FirstVariationOfAction}, as a functional by integrating the first variation times a delta function,
 \begin{align}
M[\psi]\equiv&\int_{r'}\mathcal{M}(\psi,\nabla\psi,\nabla^{2}\psi) \nonumber\\
=&\int_{r'}\Bigg(\sum_{i=a/c}iq_{i}\lambda_{i}\Lambda_{i}^{-3}e^{-iq_{i}\beta\psi(r')}\nonumber\\
-&\left(\epsilon_0+\lambda_{s}\Lambda_{s}^{-3}p^{2}\beta\frac{\sinh(u')}{u'}(\mathcal{L}^{2}(u')+\mathcal{L}'(u'))\right)\nabla^{2}\psi(r')\Bigg)\nonumber\\&\delta(r-r'),
 \end{align}
 where $u'$ denotes $p\beta \vert \nabla \phi(r') \vert$ and for clarity $\nabla\psi(r)\equiv\nabla_r\psi(r)$. Then, the second variation is
\begin{align}
\frac{\delta^{2}S}{\delta\psi(r')\delta\psi(r)}=\underbrace{\frac{\partial\mathcal{M}}{\partial\psi(r')}}_{(1)}-\underbrace{\nabla\frac{\partial\mathcal{M}}{\partial\nabla\psi(r')}}_{(2)}+\underbrace{\nabla^{2}\frac{\partial\mathcal{M}}{\partial\nabla^{2}\psi(r')}}_{(3)},
\end{align}
where
\begin{align}
(1)=&\frac{\partial\mathcal{M}}{\partial\psi(r')}=\sum_{i=a/c}q_{i}^{2}\beta\lambda_{i}\Lambda_{i}^{-3}e^{-iq_{i}\beta\psi(r')}\delta(r-r') \\
(2) =&\nabla\frac{\partial\mathcal{M}}{\partial\nabla\psi(r')}=-i\lambda_{s}\Lambda_{s}^{-3}p^{3}\beta^{2}\nabla[\nabla^{2}\psi(r')\delta(r-r') \nonumber\\
&\frac{\sinh(u')}{u'}\frac{\nabla\psi}{\abs{\nabla\psi}}(\mathcal{L}^{3}+3\mathcal{L}\mathcal{L}'+\mathcal{L}'')]\\
(3)=&\nabla_{r'}^{2}\frac{\partial\mathcal{M}}{\partial\nabla^{2}\psi(r')}=-\epsilon_0\nabla_{r'}^{2}\delta(r-r')\nonumber\\
&-i\lambda_{s}\Lambda_{s}^{-3}p^{3}\beta^{2}\nabla_{r'}\bigg[\nabla^{2}\psi\delta(r-r')\frac{\sinh(u')}{u'}\frac{\nabla\psi}{\abs{\nabla\psi}} \nonumber\\
&(\mathcal{L}^{3}+3\mathcal{L}\mathcal{L}'+\mathcal{L}'')\bigg] \nonumber\\
&-\lambda_{s}\Lambda_{s}^{-3}p^{2}\beta\nabla_{r'}\left(\frac{\sinh(u')}{u'}(\mathcal{L}^{2}+\mathcal{L}')\nabla_{r'}\delta(r-r')\right).
\end{align}
Note that we did not evaluate the last gradient derivative in $(2)$ and $(3)$ to show that the second term in $(3)$ cancels exactly with the $(2)$ term. To evaluate the second variation, at the saddle point, some useful identities can be used where one needs the properties of the Langevin function and its derivatives,
\begin{align}
&\mathcal{L}(ip\abs{\nabla\psi})=-\mathcal{L}(p\abs{\nabla\phi}), \quad
\mathcal{L}'(ip\abs{\nabla\psi})=\mathcal{L}'(p\abs{\nabla\phi}), \nonumber\\
&\mathcal{L}''(ip\abs{\nabla\psi})=-\mathcal{L}''(p\abs{\nabla\phi}).
\end{align}
Then one arrives at
\begin{align}
    \frac{\delta^{2}S}{\delta\psi(r)\delta\psi(r')} =& -\nabla_{r'}\left(\epsilon(r')\nabla_{r'}\delta(r-r')\right) \nonumber\\
    &+\sum_{i=a/c}q_{i}^{2}\beta\lambda_{i}\Lambda_{i}^{-3}e^{-iq_{i}\beta\psi(r')}\delta(r-r'). 
\end{align}
with
\begin{align}
    \epsilon(r')=\epsilon_0+\lambda_{s}\Lambda_{s}^{-3}p^{2}\beta\frac{\sinh(u')}{u'}(\mathcal{L}^{2}(u')+\mathcal{L}'(u')).
\end{align}

\subsection{Calculation of the chemical potentials in Eqs.~\eqref{ionChemicalPotential} and \eqref{solventChemicalPotential}}\label{DerivationChemicalPotentials}

Here we introduce a new three dimensional vector $x$ as an argument because the variables $r$ and $r'$ are needed for the argument of the Green's function. The particle density for the ions can be calculated as
\begin{align}
n_{i}(x)=-\frac{\partial\Gamma}{\partial v^\text{aux}_i(x)}=-\frac{\partial S[i^{-1}\phi]}{\partial v^\text{aux}_i(x)}-\frac{\upsilon}{2\beta}\text{tr}G\frac{\partial G^{-1}}{\partial v^\text{aux}_i(x)},
\end{align}
where the MF contribution is
\begin{align}
\frac{\partial S[i^{-1}\phi]}{\partial v^\text{aux}_i(x)}=-e^{\beta\mu_{i}(x)}\Lambda_{i}^{-3}e^{-q_{i}\beta\phi(x)},
\end{align}
and the 1L contribution is
\begin{align}
\frac{\partial G^{-1}(r,r')}{\partial v^\text{aux}_i(x)}=q_{i}^{2}\beta^{2}e^{\beta\mu_{i}(r')}\Lambda_{i}^{-3}e^{-q_{i}\beta\phi(r')}\delta(x-r')\delta(r-r').
\end{align}
Together this gives the particle density,
\begin{align}
&n_{i}(x)=e^{\beta\mu_{i}(x)}\Lambda_{i}^{-3}e^{-q_{i}\beta\phi(x)}\nonumber \\
&-\upsilon\frac{q_{i}^{2}\beta\Lambda_{i}^{-3}}{2}\int_{r,r'}G(r,r')e^{\beta\mu_{i}(r')}e^{-q_{i}\beta\phi(r')}\delta(x-r')\delta(r-r')\nonumber\\
&=e^{\beta\mu_{i}(x)}\Lambda_{i}^{-3}e^{-q_{i}\beta\phi(x)}-\upsilon\frac{q_{i}^{2}\beta\Lambda_{i}^{-3}}{2}e^{\beta\mu_{i}(x)}e^{-q_{i}\beta\phi(x)}G(x,x)\nonumber\\
&=\Lambda_{i}^{-3}e^{\beta\mu_{i}(x)}e^{-q_{i}\beta\phi(x)}(1-\upsilon\beta\frac{q_{i}^{2}}{2}G(x,x)).
\end{align}
Inserted into Eq.~(\ref{ComputingDensityFromPartitionFunction}) and inverted gives the chemical potential,
\begin{align}\label{chemicalPotentialIonsAppendix}
\mu_{i}=\beta^{-1}\log\frac{n_{i}(r)\Lambda_{i}^{3}}{l_{i}(r)}+q_{i}\phi(r)+v_i^\text{aux}(r)+v_i^\text{ext}(r),
\end{align}
where we define the correlation parameter,
\begin{align}\label{l_i}
l_{i}(r)\equiv1-\upsilon\beta\frac{q_{i}^{2}}{2}G(r,r),
\end{align}
A similar calculation for the solvent molecules gives
\begin{align}\label{solventChemicalPotentialAppendix}
\mu_{s}&=\beta^{-1}\log\bigg(\frac{n_{s}(r)\Lambda_{s}^{3}}{l_{s}(r)}\bigg) \nonumber\\
&-\beta^{-1}\log\bigg(\frac{\sinh(p\beta\abs{\nabla\phi(r)})}{p\beta\abs{\nabla\phi(r)}}\bigg)+v_s^\text{aux}(r)+v_s^\text{ext}(r), 
\end{align}
with
\begin{align}\label{l_s}
l_{s}(r)\equiv1+\upsilon\frac{\beta p^{2}}{2}(\mathcal{L}^{2}+\mathcal{L}')\nabla^{2}G(r,r).
\end{align}
The issue with Eq.~\eqref{chemicalPotentialIonsAppendix} and Eq.~\eqref{solventChemicalPotentialAppendix} is that the parameter \( l_j \) can potentially become negative, making the logarithm undefined. This problem arises because a reference state for measuring the chemical potential has not been established. We define a reference state chemical potential that belongs to a state of constant density $n_j^\text{ref}$ and correlation parameter $l_j^\text{ref}$, albeit with zero electrostatic potential. Adding and subtracting
\begin{align}
    \mu_j^\text{ref}=\beta^{-1}\log\bigg(\frac{n_{j}^\text{ref}\Lambda_{j}^{3}}{l_{j}^\text{ref}}\bigg)
\end{align}
in Eqs.~\eqref{chemicalPotentialIonsAppendix} and \eqref{solventChemicalPotentialAppendix} yields Eqs.~\eqref{ionChemicalPotential} and \eqref{solventChemicalPotential}, where only $l_j/l_j^\text{ref}$ appears in the argument of the logarithm.
\subsection{The local-density-approximation for the correlation parameters Eqs.~\eqref{ScalingParameterIon} and \eqref{ScalingParameterSolvent}}\label{Appendix:Local-density}
Due to the anisotropy of the metal-electrolyte interface, the Green's function depends not only on the distance, $\abs{r-r'}$, but explicitly on the two vectors $r$ and $r'$.\cite{netzElectrostatisticsCounterionsPlanar2001, markovichIonicProfilesClose2016} This work opts for a local-density-approximation (LDA), which neglects the spatial dependence of the fields in the differential equation, Eq.~\eqref{GreenFunctionCanonical} to obtain an analytical result for $G(r-r)$. Only after obtaining $G(r-r)$ is the spatial dependence of the fields restored. This approach is similar to the LDA of the exchange-correlation functional in DFT, where the inhomogeneous gas functional is derived assuming a uniform electron gas.\cite{hohenbergInhomogeneousElectronGas1964,kohnSelfConsistentEquationsIncluding1965} 

The Green's function , Eq.~\eqref{GreenFunctionCanonical}, in LDA is given by Eq.~\eqref{DiffEqGreen}.  The equal-point Green's function and its Laplacian can be obtained by Fourier transforming Eq.~(\ref{DiffEqGreen}) and defining
\begin{align}\label{DerivationGreenFunction}
G(r,r)&=\int_{\abs k<k_{max}}\frac{d^{3}k}{(2\pi)^{3}}\frac{1}{\epsilon(r)}\cdot\frac{1}{\boldsymbol{k}^{2}+\lambda_{D}^{-2}}\\
&=\frac{1}{2\pi^{2}\epsilon(r)}\left(\frac{2\pi}{a_{i}}-\frac{1}{\lambda_{D}}\arctan\left(2\pi\frac{\lambda_{D}}{a_{i}}\right)\right),
\end{align}
and
\begin{align}\label{DerivationGreenFunction2}
&\nabla^{2}G(r,r)=-\int_{\abs k<k_{max}}\frac{d^{3}k}{(2\pi)^{3}}\boldsymbol{k}^{2}\tilde{G}(k)\\
&=-\frac{1}{2\pi^{2}\epsilon(r)}\left(\frac{8\pi^{3}}{3a_{s}^{3}}-\frac{2\pi}{\lambda_{D}^{2}a_{s}}+\frac{1}{\lambda_{D}^{3}}\arctan\left(2\pi\frac{\lambda_{D}}{a_{s}}\right)\right),
\end{align}
where a maximum wavelength cutoff $k_{max}$ was introduced to fix the divergence of the Green's function at the same argument. Crucially, the cutoff for $G$ ($a_i$) is different from the cutoff of $\nabla^2G$ ($a_s$). This is necessary in order to reproduce experimental data of dielectric permittivity and activity coefficient, \textit{cf.} Sec. \ref{SubSec:Bulk}.  Inserting the results into the correlation parameters, Eqs.~(\ref{l_i}) and (\ref{l_s}) give Eqs.~\eqref{ScalingParameterIon} and \eqref{ScalingParameterSolvent}).

\subsection{Calculation of the functional derivatives in section \ref{SubSec:FunctionalDerivative}}\label{App:FunctionalDerivatives}
In this subsection, we discuss the computation of the variational derivatives of the LDA correlation functional, Eq.~\eqref{CorrelationFunctional}, with respect to $\phi$ and $n_j$. 

\subsubsection{Functional derivative of the 1L-LDA functional w.r.t. $\phi$ in Eq.~\eqref{LDAVariationalDerivative1}}
In evaluating the functional derivative with respect to $\phi$, it is important to recognize that the parameters $l_j$ themselves depend on $\nabla\phi$,
\begin{align}
&\frac{\partial l_{i}}{\partial\nabla\phi} = \frac{3}{2}\frac{p^{3}\beta^{2}n_{s}}{\epsilon(r)\abs{\nabla\phi}}(2\mathcal{L}\mathcal{L}'+\mathcal{L}'')(1-l_{i})\nabla\phi, \\
&\frac{\partial l_{s}}{\partial\nabla\phi} = \frac{\beta p}{\abs{\nabla\phi}} \frac{(2\mathcal{L}\mathcal{L}'+\mathcal{L}'')}{(\mathcal{L}^{2}+\mathcal{L}')}(l_s-1)\nabla\phi.
\end{align}
This enables us to compute 
\begin{align}\label{ComputFunctionalDerivative1}
     \frac{\delta \mathcal{F}^\text{corr,$1$L-LDA}_\text{sol}}{\delta\phi(r)} &= \frac{\delta}{\delta\phi(r)}\int_r\sum_j n_j(r)\epsilon_j^\text{corr}(r)  \nonumber\\&+ \frac{\delta}{\delta\phi(r)}\frac{\upsilon}{2\beta}\text{tr}\log\beta G^{-1}[\phi(r)].
\end{align}
The first part on the r.h.s gives
\begin{align}
    \frac{\delta}{\delta\phi(r)}&\int_{r'}\sum_{j}n_{j}(r')\epsilon_{j}^{\text{corr}}(r') \nonumber\\
    &= \nabla\Big[-\frac{3}{2}\frac{p^{3}\beta n_{s}}{\epsilon(r)\abs{\nabla\phi}}(2\mathcal{L}\mathcal{L}'+\mathcal{L}'')\sum_{i}n_{i}\frac{(1-l_{i})^{2}}{l_{i}^{2}} + \nonumber \\
    &\frac{n_{s}p}{\abs{\nabla\phi}}\cdot\frac{(2\mathcal{L}\mathcal{L}'+\mathcal{L}'')}{(\mathcal{L}^{2}+\mathcal{L}')}\frac{(1-l_{s})^{2}}{l_{s}^{2}})\nabla\phi\bigg].
\end{align}
Note that the two terms $\sim \frac{(1-l_j)^2}{l_j^2}$ are both $\sim \mathcal{O}(\upsilon^{2})$ and can therefore be neglected. Both terms result from the explicit dependence of the correlation energy, $\epsilon^\text{corr}_j$, on the electric field. To compute the second part on the r.h.s. of Eq.~\eqref{ComputFunctionalDerivative1}, we need the variational derivative of the inverse Green's function. For that purpose, we rewrite the Green's function as a functional,
\begin{align}
G^{-1}(r,r')=\int_{x}dx\;\mathcal{M}(x,r,r'),
\end{align}
with 
\begin{align}
&\mathcal{M}(x,r,r')=\bigg(\sum_{i=a/c}q_{i}^{2}\beta n_{i}(x)\delta(r-x) \nonumber\\
&-p^{3}\beta^{2}n_{s}\frac{\nabla\phi}{\abs{\nabla\phi}}\nabla^{2}\phi\left(\mathcal{L}^{3}+\mathcal{L}\mathcal{L}'+2\mathcal{L}\mathcal{L}'+\mathcal{L}''\right)\nabla_{x}\delta(r-x) \nonumber\\
&-\epsilon(x)\nabla_{x}^{2}\delta(r-x)\bigg)\delta(x-r'),
\end{align}
which lets us write the functional derivative as
\begin{align}
\frac{\delta G^{-1}}{\delta\phi(x)}&=\frac{\partial\mathcal{M}}{\partial\phi(x)}-\nabla\frac{\partial\mathcal{M}}{\partial\nabla\phi(x)}+\nabla^{2}\frac{\partial\mathcal{M}}{\partial\nabla^{2}\phi(x)},
\end{align}
which yields
\begin{align}
\frac{\delta G^{-1}}{\delta\phi(x)}=&-\nabla\Big(\frac{p^{3}\beta^{2}n_{s}\nabla\phi(x)}{\abs{\nabla\phi(x)}} \nonumber\\
&\left(\mathcal{L}^{3}+\mathcal{L}\mathcal{L}'+2\mathcal{L}\mathcal{L}'+\mathcal{L}''\right) \nonumber\\
&\nabla\delta(r-x)\nabla\delta(x-r')\Big).
\end{align}
The trace is then simply obtained,
\begin{align}
\text{tr}G\frac{\delta G^{-1}}{\delta\phi(x)}=\nabla\bigg[2\beta\upsilon^{-1}\frac{n_{s}p\nabla\phi(x)}{\abs{\nabla\phi}}\frac{(\mathcal{L}^{3}+3\mathcal{L}\mathcal{L}'+\mathcal{L}'')}{\mathcal{L}^{2}+\mathcal{L}'}(1-l_{s})\bigg],
\end{align}
and contains a correction $\mathcal{O}(\upsilon)$. Plugging in the result, we arrive at Eq.~\eqref{LDAVariationalDerivative1}. 
\subsubsection{Functional derivative of the 1L-LDA functional w.r.t. $n_j$ in Eq.~\eqref{LDAVariationalDerivative2}}
Goal of this section is to calculate the functional derivative,
\begin{align}
    \frac{\delta \mathcal{F}^\text{corr,$1$L-LDA}_\text{sol}}{\delta n_j(r)} &=\frac{\delta}{\delta n_j(r)}\int_r\sum_j n_j(r)\epsilon_j^\text{corr}(r)   \nonumber\\
    &+ \frac{\delta}{\delta n_j(r)}\frac{\upsilon}{2\beta}\text{tr}\log\beta G^{-1}[\phi(r)]
\end{align}
The functional derivative with respect to $n_j$ is complicated due to the dependence of $l_j$ on $n_j$. However, we will see that the derivatives of $l_j$ and the $\text{tr $\log$}$ term will be only of sub-leading order and can therefore be neglected. 

The derivative in the first part yields
\begin{align}
\frac{\delta}{\delta n_j(r)}\int_r\sum_j n_j(r)\epsilon_j^\text{corr}(r)&=-\beta^{-1}\log(l_{i}/l_i^\text{ref}) \nonumber \\
&+\beta^{-1} \Big(\frac{l_{i}-1}{l_{i}}-\frac{1}{2}\frac{n_{i}q_{i}^{2}}{\sum_{i}q_{i}^{2}n_{i}}\frac{(l_{i}-1)^{2}}{l_{i}^{2}}\Big) \nonumber\\
&-\beta^{-1}\frac{3}{2}\frac{n_{s}q_{i}^{2}}{\sum_{i}q_{i}^{2}n_{i}}\frac{(1-l_{s})^{2}}{l_{s}^{2}},
\end{align}
whereas the derivative of the  $\text{tr $\log$}$  term can be obtained with
\begin{align}
\frac{\delta G^{-1}(r,r')}{\delta n_{i}(x)}=q_{i}^{2}\beta\delta(r-x)\delta(x-r'),
\end{align}
thus
\begin{align}
\frac{\upsilon}{2\beta}\text{tr}G\frac{\delta G^{-1}}{\delta n_{i}(r)}&=\frac{\upsilon}{2\beta}q_{i}^{2}\beta G(r,r) \nonumber\\&=\beta^{-1}(1-l_{i}),
\end{align}
which can be summarized to,

\begin{align}
 \frac{\delta \mathcal{F}^\text{corr,$1$L-LDA}_\text{sol}}{\delta n_j(r)}&=-\beta^{-1}\log(l_{i}/l_i^\text{ref})\nonumber\\
&-\beta^{-1}\big(\frac{(l_{i}-1)^{2}}{l_{i}} \nonumber\\
&+\frac{1}{2}\frac{n_{i}q_{i}^{2}}{\sum_{i}q_{i}^{2}n_{i}}\frac{(l_{i}-1)^{2}}{l_{i}^{2}}\nonumber\\
&+\frac{3}{2}\frac{n_{s}q_{i}^{2}}{\sum_{i}q_{i}^{2}n_{i}}\frac{(1-l_{s})^{2}}{l_{s}^{2}}\big).
\end{align}
Evidently, if neglecting the term $\mathcal{O}(\upsilon^2)$, one arrives at the result Eq.~\eqref{LDAVariationalDerivative2}. 

\bibliographystyle{apsrev4-1}

\end{document}